\documentclass[acmsmall]{acmart}
\usepackage{tabularx}
\usepackage{xcolor}
\AtBeginDocument{%
	\providecommand\BibTeX{{%
			\normalfont B\kern-0.5em{\scshape i\kern-0.25em b}\kern-0.8em\TeX}}}

\setcopyright{acmcopyright}
\acmJournal{CSUR}
\acmYear{2022} 
\acmVolume{1} 
\acmNumber{1} 
\acmArticle{1} 
\acmMonth{1} 
\acmPrice{15.00}
\acmDOI{10.1145/3511094}

\begin{document}
	
	
	\title{Energy Efficient Computing Systems: Architectures, Abstractions and Modeling to Techniques and Standards}

	\author{Rajeev Muralidhar}
	\email{rajeev.muralidhar@student.unimelb.edu.au, rajeevm@ieee.org}
	\affiliation{
		\institution{University of Melbourne and Amazon Web Services}
		\city{Parkville}
		\state{Victoria}
		\postcode{3010}
		\country{Australia}
	}
	
	\author{Renata Borovica-Gajic}
	\email{renata.borovica@unimelb.edu.au}
	\affiliation{
		\institution{University of Melbourne}
		\city{Parkville}
		\state{Victoria}
		\postcode{3010}
		\country{Australia}
	}
	
	\author{Rajkumar Buyya}
	\email{rbuyya@unimelb.edu.au}
	\affiliation{
		\institution{University of Melbourne}
		\city{Parkville}
		\state{Victoria}
		\postcode{3010}
		\country{Australia}
	}
	
	
	
	
	
	
	
	
	\renewcommand{\shortauthors}{Rajeev, et al.}
	
	\begin{abstract}
		Computing systems have undergone a tremendous change in the last few decades with several inflexion points. While Moore's law guided the semiconductor industry to cram more and more transistors and logic into the same volume, the limits of instruction-level parallelism (ILP) and the end of Dennard's scaling drove the industry towards multi-core chips. More recently, we have entered the era of domain-specific architectures and chips for new workloads like artificial intelligence (AI) and machine learning (ML). These trends continue, arguably with other limits, along with challenges imposed by tighter integration, extreme form factors and increasingly diverse workloads, making systems more complex to architect, design, implement and optimize from an energy efficiency perspective. Energy efficiency has now become a first order design parameter and constraint across the entire spectrum of computing devices. \\
		Many research surveys have gone into different aspects of energy efficiency techniques implemented in hardware and microarchitecture across devices, servers, HPC/cloud, data center systems along with improved software, algorithms, frameworks, and modeling energy/thermals. Somewhat in parallel, the semiconductor industry has developed techniques and standards around specification, modeling/simulation, benchmarking and verification of complex chips; these areas have not been addressed in detail by previous research surveys. This survey aims to bring these domains holistically together, present the latest in each of these areas, highlight potential gaps and challenges, and discuss opportunities for the next generation of energy efficient systems. The survey is composed of a systematic categorization of key aspects of building energy efficient  systems - (1) \textit{specification} - the ability to precisely specify the power intent, attributes or properties at different layers (2) \textit{modeling} and \textit{simulation} of the entire system or subsystem (hardware or software or both) so as to be able to experiment with possible options and perform what-if analysis, (3) \textit{techniques} used for implementing energy efficiency at different levels of the stack, (4) \textit{verification} techniques used to provide guarantees that the functionality of complex designs are preserved, and (5) \textit {energy efficiency benchmarks, standards and consortiums} that aim to standardize different aspects of energy efficiency, including cross-layer optimizations. 
	\end{abstract}

	
	

	\maketitle

	\section{Introduction}
	\label{section:introduction}
	
\begin{figure}[ht]
		\centering
		\includegraphics[width=\linewidth]{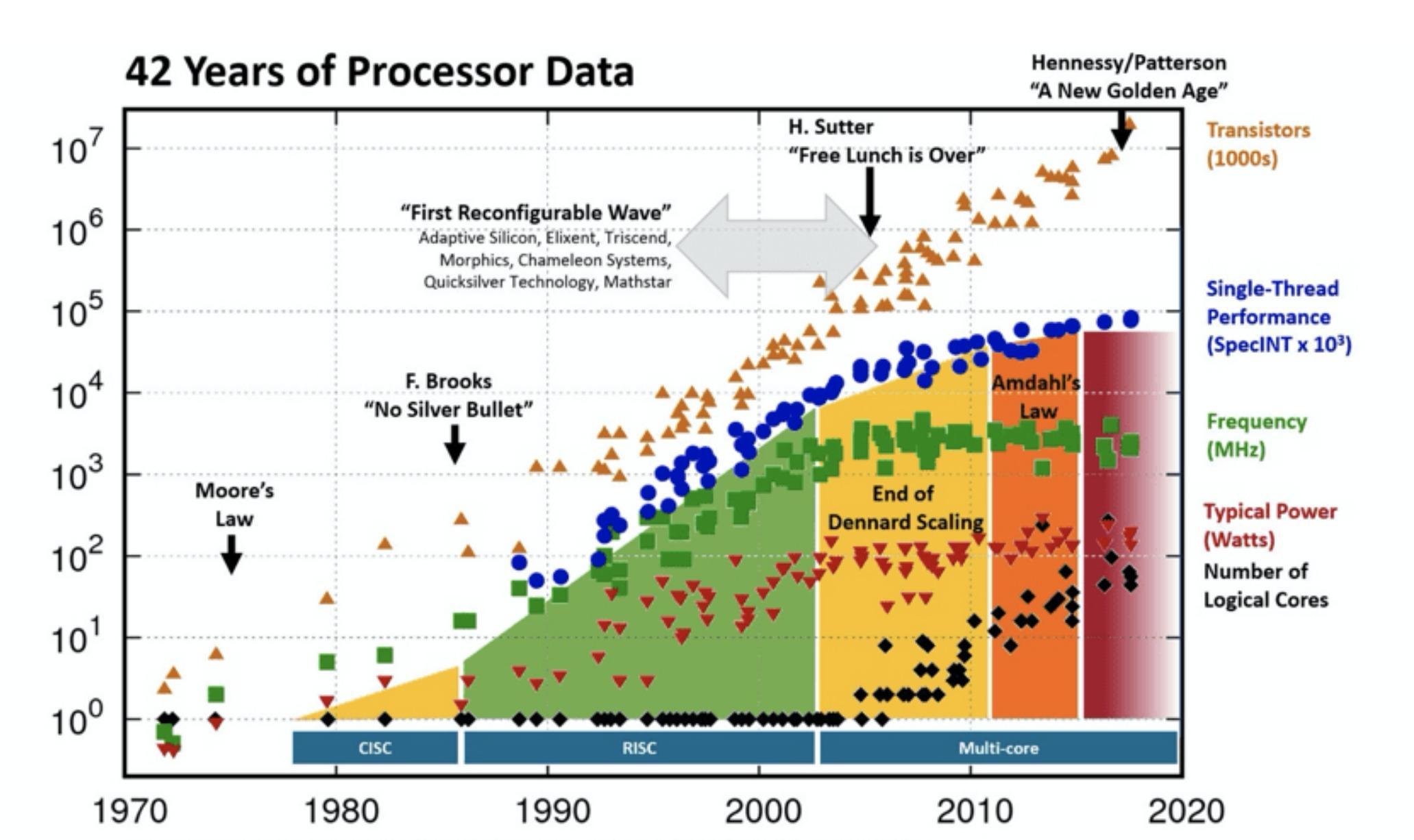}
		\caption{Microprocessor trend data during 1972-2020. Hennessey and Patterson, Turing Lecture 2018 \cite{henessey-patterson-golden-age-comp-arch}, overlaid with "42 Years of Processors Data" \cite{karl-rupp-42-years-trend}}
		\label{fig:42-years-microprocessor-trends}
		\Description{Microprocessor trend data during 1972-2020}
	\end{figure}

	The computing industry has gone through tremendous change in the last few decades. While Moore's law \cite{moore:1965} drove the semiconductor industry to cram more and more transistors and logic into the same volume, the end of Dennard's scaling \cite{dennard74design} limited how much we could shrink voltage and current without losing predictability, and the Instruction Level Parallelism (ILP) wall (David Wall et al. \cite{davidwall-ilplimits}) defined the start of the multi-core and tera-scale era \cite{intelterascale}. As the number of cores and threads-per-core increased, energy efficiency and thermal management presented unique challenges. We soon ran out of parallelizability as well, both due to limits imposed by Amdahl's law \cite{Amdahl67} and a fundamental lack of general purpose parallelizable applications and workloads. Fig \ref{fig:42-years-microprocessor-trends}, referenced from Karl Rupp \cite{karl-rupp-42-years-trend} shows 42 years of microprocessor trends taking into account transistor density, performance, frequency, typical power and number of cores. The figure is based on known transistor counts published by Intel, AMD and IBM's Power processors and it also overlays the key architectural inflexion points detailed by Henessey and Patterson in  \cite{henessey-patterson-golden-age-comp-arch}. The graph, as well as studies such as Fagas et al. \cite{Fagas17}, illustrate that as transistor count and power consumption continues to increase, frequency and the number of logical cores has tapered out. Furthermore, as Moore's Law slows down, while energy efficiency has improved, power density continues to raise across the spectrum of computing devices (Mack et al. \cite{50-years-moores-law}). With multi-core architectures reaching its limits, the last few years have seen the emergence of domain specific architectures to attain the best performance-cost-energy tradeoffs for well defined tasks. Systems also evolved from multi-chip packages to system-on-a-chip (SOC) architectures with accelerators like Graphics Processing Units (GPU), imaging, Artificial Intelligence (AI)/deep learning and networking, integrated with high-bandwidth interconnects. Workloads such as deep learning require massive amounts of data transfer to/from memory, leading to the \textit{memory wall}, which is the bottleneck imposed by the bandwidth of the channel between the CPU and memory subsystems. Recent memory technologies like Non-Volatile Random Access Memory (NVRAM), Intel's Optane, Spin Transfer Torque RAM (STT-SRAM), and interfaces such as Hybrid Memory Cube (HMC) \cite{hmc1} and High Bandwidth Memory (HBM) \cite{lee20151} that enable high-performance RAM interfaces have pushed the boundaries of the memory wall. Standards such as PCIe \cite{pci-sig} and the more recent Compute Express Link (CXL) \cite{cxl} are industry standards for integrating accelerators, memory and compute elements. Deep learning has also triggered looking at the traditional von-Neumann architectural model and its limits thereof and several non-von Neumann models have now gained popularity, such as those based on dataflow, spiking neural networks, neuromorphic computing and bio-inspired computing (Ganguly et al. \cite{Ganguly2019TowardsEE}).


	
    The nature of computing systems has transformed across the spectrum of devices, from being pure compute-based to being a mixture of CPUs, GPUs, accelerators and Field Programmable Gate Arrays (FPGA). Heterogeneous computing capabilities are now also available on "edge devices" such as the Raspberry PI, Google's Coral Tensor Processing Unit \cite{googletpu} and Intel's Movidius \cite{intelmovidiusmyriad2}. As devices have shrunk, the industry is struggling to eliminate the effects of thermodynamic fluctuations, which are unavoidable at lower technology nodes (sub-10nm scale). Even as architectures become more energy efficient, recent research has shown that workloads such as deep learning consumes significant energy (Schwartz et al. \cite{greenai}). Ironically, deep learning was inspired by the human brain, which is remarkably energy efficient. Shrinking and extreme form factors, diverse workloads and computing models have thus greatly accelerated the limitations imposed by fundamental physics and architectural, power and thermal walls. 
	    
    Designing energy efficient systems present unique challenges due to the domain-specific processing capabilities required, heterogeneous nature (workloads that can run on CPUs, GPUs or specialized chips), system architecture (high bandwidth interconnects for the enormous amounts of data transfer required) and extreme form factors (with devices capable of doing Tiny ML, which is the ability to do machine learning in less than 1 mW of power \cite{tinyml}). Systems have become complex to architect, design, implement and verify, with energy efficiency transforming into a multi-disciplinary art requiring expertise across hardware/circuits, process technology, microarchitecture, domain-specific hardware/software, firmware/micro-kernels, operating systems, schedulers, thermal management, virtualization and workloads, only to name a few. While specific end systems (IoT, wearables, servers, HPC) need some techniques more aggressively than others due to the constraints, the underlying energy efficiency techniques tend to overlap across systems and hence we need to take a holistic view as we look to improve and architect next generation systems.

	\subsection{Related Surveys}
	\label{subsection:related-surveys}
	Several research surveys have looked at energy efficiency techniques used in hardware, circuits/RTL, microarchitecture and process technology, across the spectrum of computing systems. Another area of active research has been around modeling and simulation of power, performance and thermals for individual hardware components (processors, memory, GPUs, and accelerators), system-on-a-chip (SOC) and the entire system. In parallel, techniques and standards have evolved in the semiconductor and Electronic Design and Automation (EDA) industry around specification and verification of large, complex chips. The industry has also collaborated to build highly optimized software/system level techniques and has defined energy related benchmarks, regulations and standards. This survey brings the domains together and presents the latest in each area, highlights potential gaps/challenges, and discusses opportunities for next generation energy efficient systems. Some current related research surveys are listed in Table \ref{tab:ee-related-survey-list} - this list is, by no means exhaustive, but merely points to some key surveys or books in respective areas.
	

	\begin{table}
		\caption{Summary of Energy Efficiency Related Surveys}
		\label{tab:ee-related-survey-list}
		\begin{tabular}{ccl}
			\toprule
			Topic & Key survey or book \\
			\midrule
			Energy efficiency/sustainability, metrics in cloud & 
			Mastelic et al. \cite{cloud-ee-survey}, Gill et al. \cite{buyya-sustainable-computing} \\
			Energy efficiency techniques in hardware, circuits & 
			Venkatachalam et al. \cite{ee-microprocessors} \\
			Hardware techniques for energy efficiency in CPUs, GPUs & 
			Mittal et al. \cite{mittal-vetter-ee-gpu} \\
			Energy Efficiency of compute nodes &  
			Kaxiras and Martonosi \cite{martonosi-book-comp-arch-energy-eff}\\
			Energy efficiency at data center level & 
			Barroso and Hoelzle \cite{hoelzle-barroso-datacenter} \\
			\bottomrule
		\end{tabular}
	\end{table}
	
	\subsection{Need for a holistic approach to energy efficiency}
	\label{subsection:need-holistic-approach}
	Designing energy efficient systems is now a virtuous cycle and cannot be done in hardware or software alone, or in isolation of other domains or components due to diverse architectures, hardware/software interactions and varied form factors. Power-related constraints have to be imposed through the entire design cycle in order to maximize performance and reliability. In the context of large and complex chip designs, reliability and minimizing power dissipation have become major challenges for design teams, which have dependencies on software as well. Creating optimal low-power designs involves making trade-offs such as timing-versus-power and area-versus-power at different stages of the design flow. Additionally, trade-offs that are applied at a certain phase of the chip have implications on future software techniques that push the boundaries of what the chip has been designed to do. In many cases, if certain design choices are known ahead of time, specific workloads will benefit from them with respect to energy efficiency. 
	
\begin{figure}[ht]
		\centering
		\includegraphics[width=\linewidth]{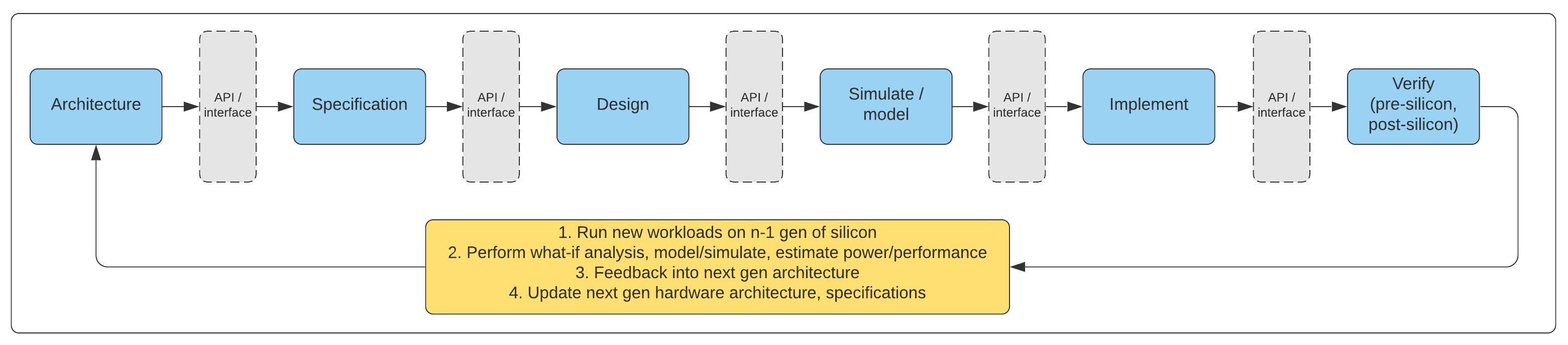}
		\caption{Phases of energy efficient system design}
		\label{fig:ee-system-design-phases}
		\Description{Phases of energy efficient system design}
	\end{figure}

	Feedback from running real workloads on current generation systems is used in architecting next generation systems. Architects need to perform "what-if" analysis using different algorithmic knobs at different stages as illustrated in Figure \ref{fig:ee-system-design-phases}. For example, it is important to simulate different techniques of frequency state selection, their transition latencies and the impact of these states on different workloads. Adding or removing power efficiency features can make or break the chip launch timeline, which could have market implications and could impact the company's future itself. The ability to model power consumption of different hardware components across generations of hardware in a standardized manner has become a key focus of industry efforts such as the IEEE P2416 standard for power modeling \cite{ieeep2416}. As another example, the ability to run a real workload on a simulated future design and making use of new power/performance features is an important to expose bugs in the underlying hardware. If these bugs are found later in post-silicon, it could cause unacceptable delays due to a hardware re-spin. Such scenarios need information exchange across layers of the hardware-software stack - such as new frequency states being exposed, how the OS and higher layers can make use of it and the ability to model performance gain therein. The goal of the recent IEEE P2415 \cite{ieeep2415} is to build cross layer abstractions such as this to facilitate easier information exchange across different layers of the stack as well as different phases of architecture, modeling and verification.


	
	Energy efficiency in HPC systems has also become important of late. The Energy Efficient HPC (EEHPC) \cite{eehpc} Consortium is a group focused on driving implementation of cross layer energy conservation measures and energy efficient designs HPC systems. The working groups cover several aspects of energy efficient HPC - infrastructure (cooling, highly efficient power sources), algorithms and runtime (energy and power aware job scheduling), and specifications (Power API). Similarly, the Global Extensible Open Power Manager (GEOPM) (Eastep et al. \cite{geopm-hpc}) is an open source runtime HPC framework for enabling new energy management strategies at the node, cluster and data center level.
	
	Holistic energy efficiency across layers and across phases of evolution is crucial and cannot ignore any of the platform components; neither can it be done in hardware or software alone and must encompass all aspects of energy efficient system design - from architecture to modeling/simulating to implementing and optimizing each component as well as the system as a whole. 
	
	\subsection{Contributions of This Survey}
	\label{subsection:contributions}
	
	Previous surveys have looked at energy efficiency in hardware/microarchitecture, at different layers (software and algorithms) and at different systems (devices, servers, cluster and cloud). In such surveys, it is assumed that hardware architectures and features of energy efficiency in hardware evolve on their own, and software then takes the best possible approach by designing energy aware algorithms. Additionally, several industry trends, benchmarks, standards and consortiums related to energy efficiency have not been surveyed in detail. As systems become complex, energy efficiency considerations must be imposed across the entire cycle - from hardware/system architecture, design, specification, modeling/simulation, to higher layers of software algorithms that use these features to optimize the system. With that goal in mind, this survey is composed of a systematic categorization of the following energy efficiency methods across the wide spectrum of computing systems:
	\begin{enumerate}
		\item \textit{Energy Efficiency Techniques}: This could be at different levels of the hierarchy - circuit/RTL, microarchitecture, CPU, GPU or other accelerators, and/or at software/system level.
		\item \textit{Specification} of the energy efficiency technique: This involves specifying the technique in a standardized manner, and includes cross-layer abstractions and interfaces (hardware, hardware-firmware, firmware-OS, and OS-applications). 
		\item \textit{Modeling and Simulation}: Given a set of techniques for energy efficiency, this involves modeling/simulating the functionality/technique of the component or set of components, and run real workloads (or traces of a real workload).
		\item  \textit{Verification}: Given each of the above, this involves verifying the energy efficiency of the entire system with different thermal constraints, real workloads and different form factors.
		\item \textit{Energy Efficiency Benchmarks, Standards and Consortiums}: Recent trends at standardizing different aspects of energy efficiency at IEEE and other industry consortiums is an important area of research/industrial collaboration.
	\end{enumerate}
	
	
	
	
	\subsection{Organization of this Paper}
	\label{subsection:survey-organization}
    The rest of the paper is organised as follows:
	\begin{enumerate}
		\item In Section \ref{section:power-thermals-basics}, we discuss the basics of power and energy, thermal dissipation, and fundamental techniques used to build energy efficient systems.

		\item Section \ref{section:archtrends} elaborates on recent architectural inflexion points, evolution of energy efficiency features and upcoming trends.
		
		\item Section \ref{section:microarchitectural-techniques} discusses microarchitectural techniques used in CPUs, GPUs, memory and domain-specific accelerators.
		
		\item Section \ref{section:specification} discusses \textit{specification} of power management techniques. Being able to capture the power intent in a formal description is key to design, modeling/simulation as well as verification of the system as a whole. This is also fundamental to the Electronic Design Automation (EDA) industry, IP-reuse and building complex ssytems. We survey specifications and abstractions at different levels of the hierarchy.
		
		\item Section \ref{section:modeling-simulation} covers \textit{modeling and simulation} of power, performance and thermal dissipation across processors, GPUs, accelerators, SOC and complete systems. We describe some state of the art modeling and simulation tools and technologies in use today.
				
		\item In Section \ref{section:system-level-techniques}, we cover \textit{system and software techniques} used for energy efficiency. In this, we cover energy efficiency techniques implemented in firmware, device drivers, and operating systems such as Linux and Windows. Where relevant, we also provide methods used across different classes of designs, like mobile processors, data centers, servers, etc.
		
		\item Section \ref{subsection:recent-system-level-pm} discusses recent advances in system level energy efficiency implemented across real products from Intel, AMD, ARM, including the emergence of custom-built high performance ARM designs such as AWS Graviton3, and ARM in HPC. 
		
		\item Section \ref{section:verification} covers \textit{verification} of power management design and techniques in large SOCs and systems. 
		
		\item In Section \ref{section:energy-standards}, we survey \textit{energy efficiency related benchmarks, standards and consortiums} that are trying to address energy efficiency through regulations, standardization of abstractions, energy/performance models and cross-layer optimizations.
		
		\item In Section \ref{section:discussion}, we will discuss the road ahead for next generation of energy efficient systems and in Section \ref{section:summary}, we offer our summary and conclusions. 
		
	\end{enumerate}
	
	

\section{Basics of power/thermal dissipation, energy efficiency techniques}
	\label{section:power-thermals-basics}
	
	This section provides a very brief background of the basics of power dissipation and how process technologies impact power. The different components of power dissipation in CMOS devices are exhaustively covered in Kaxiras and Martonosi \cite{martonosi-book-comp-arch-energy-eff}. A short summary is provided here. 
	
	\subsection{Energy Metrics}
	\label{subsec:energy-metrics}
	
	\textit{Energy}: Energy, measured in Joules, is one of the most fundamental metrics, and is of wide interest across all kinds of computing systems today, but especially so in mobile, wearable or IoT platforms where energy usage relates closely to battery lifetime. This metric is now of significant importance in larger systems as well. Energy consumption ranks as one of the leading operating costs across the world today and thus reducing energy usage is crucial for all kinds of computing systems.
	
    \textit{Power}: Power is the rate of energy dissipation. The unit of power is watts (W), which is Joules per second. A related metric, \textit{power density}, is power per unit area. This is useful for thermal studies and optimization as power spread over a smaller area can be quite challenging, especially without active cooling, which is the case with most IoT, wearables or smartphones.

    Another useful metric is \textit{Energy-per-Instruction (EPI)}, which is used more in the architectural context. It describes the energy required to execute an instruction. 
    
    \textit{Energy-Delay Product (EDP)} is an important metric that combines energy and performance and is used for measuring overall energy-performance of workloads. Since EDP is a product of energy consumed and the time taken to execute a workload, if either energy or  delay increase, the EDP will increase. Hence, lower EDP is desirable. 
    
	\subsection{Power Consumption and Dissipation}
	\label{subsection:power-basics}
	The primary sources of power dissipation in CMOS devices are:
	\begin{enumerate}
	    \item Switching power or dynamic power
	    \item Leakage power
	\end{enumerate}
	There are other sources as well - short circuit power and static power, however we will stick to these two major sources as they relate the most to energy efficiency techniques discussed in this paper. The book by Kaxiras and Martonosi \cite{martonosi-book-comp-arch-energy-eff} discusses other sources of power dissipation in a lot more detail.
	
	Thus, the average power is equal to the sum of the dynamic and leakage power. 
	
	$$
    {\rm P}_{{\rm avg}}= {\rm P}_{{\rm dynamic}} + {\rm P}_{{\rm leakage}} \\
    $$
    $$
    {\rm P}_{{\rm dynamic}} = A C V^2 f \\
    $$
    The first term \textit{dynamic power}, or \textit{switching power}, depends on supply voltage \textit{V}, clock frequency \textit{f}, node capacitance \textit{C} (which in turn, depends on wire lengths), and switching activity factor \textit{A} (how frequently wires transition from 0 to 1, or from 1 to 0). Dynamic power can be lowered by reducing switching activity and clock frequency, which affects performance; and also by reducing capacitance and supply voltage.
    
    \textit{Leakage power} is a function of the supply voltage V, the switching threshold voltage, temperature and the transistor size. While dynamic power is dissipated only when switching, leakage power is continuous. Of the different leakage components (reverse bias current, gate oxide leakage, etc.), \textit{sub-threshold leakage power} is the most dominant one and it represents the power dissipated by a transistor whose gate is intended to be off. The main reason behind this leakage is that transistors do not have ideal switching characteristics, and thereby leak a non-zero amount of current even for voltages lower than the threshold voltage. In smaller geometries, leakage power has become the dominant consumer of power. Leakage energy now represents 20–40\% of the power budget of microprocessors in current and near-future fabrication technologies. 
    Techniques such as LTEC (Low Temperature Effect Compensation) are used to compensate for low/high temperature scenarios \cite{intel/ltec}. 
	
	Techniques such as \textit{clock gating} are used to save energy by reducing activity factors during a hardware units idle periods. The clock frequency \textit{f}, in addition to influencing power dissipation, also influences the supply voltage. Typically, higher clock frequencies will mean maintaining a higher supply voltage. Thus, the combined \textit{(V, f)} portion of the dynamic power equation has a cubic impact on power dissipation. Strategies such as dynamic voltage and frequency scaling (DVFS) try to exploit this relationship to reduce power accordingly.

	\subsection{Thermal Dissipation}
	\label{subsection:thermal-basics}
    Thermal behavior depends on power dissipation and density, since temperature is essentially a function of how much power is dissipated in a region versus how that region is cooled. On the other hand, power also depends on temperature. As described in Kaxiras and Martonosi \cite{martonosi-book-comp-arch-energy-eff}, thermal models can be built on analogies between heat transfer and electrical phenomena. Power dissipation results in heat, and this heat flows through regions based on their thermal resistance (R). The amount of heat flow can be analogized to current (I), and the heat difference between two regions on a chip is analogous to voltage (V ). Because there are time dependencies in both the power dissipation and in its relationship to heat flow and thermal impedance, a capacitance (C) is also modeled. Thus, time-dependent RC models remain the best way to model localized thermal behavior on a chip. These models are used heavily in building complex thermal models of complete SOCs and form factors.
    
    \subsection{Basic Energy Efficiency Techniques}
	\label{subsection:basic-pm-techniques}
	
	The basic dynamic power equation 
	$$
	{\rm P}_{{\rm dynamic}} = {A C  V^2  f} 
	$$
	clearly shows that power consumption can be reduced by reducing the activity factor \textit{(A)}, supply voltage \textit{(V)}, and operating frequency \textit{(f)}. The fundamental techniques that are used to accomplish this are \textit{clock gating},  \textit{power gating} and \textit{dynamic voltage frequency scaling (DVFS)}. 
	
	\subsubsection{Clock and Power Gating}
	\label{subsubsection:clock-power-gating}
	
	\textit{Clock gating} works by removing the clock signal when the circuit is not in use, at the cost of adding more logic to a circuit. The key idea is to disable portions of the circuitry so that the flip-flops in them do not have to switch states (which contributes to dynamic power). When not being switched, the switching power consumption goes to zero, and only leakage currents are incurred. The goal here is to reduce or eliminate excess activity that does not have any effect on the computation being performed. This activity could be at any granularity - from the tiniest circuits and individual flip-flops, to whole functional units, or even larger structures and whole subsystems (for example, memory, I/O, CPU). 

	\textit{Power Gating}, or gated Vdd approach, is used for larger functional units. In this, the voltage to the functional unit is shut off. The key idea is as follows: power gating is achieved by using a suitably sized header or footer transistor for a circuit block that is deemed to be a power-gating candidate. When the logic detects the onset of a sufficiently long idle period of the target circuit block, a “sleep” signal is applied to the gate of the header or footer transistor to turn-off the supply voltage to the circuit block. Similarly, once it is determined that the circuit block is being requested for use, the “sleep” signal is de-asserted to restore the voltage at the virtual Vdd. Hu et al. \cite{powergating-microarch} describes several different techniques used in microarchitecture.
	
	\subsubsection{Dynamic Power Management}
	\label{subsubsection:dynamic-pm}
	Largely, system level techniques employ either \textit{race-to-halt} or \textit{crawl-to-idle} philosophy. \textit{Race-to-halt} attempts to reduce dynamic power consumption by proposing that the highest frequency is used to complete the task as fast as possible, and once finished, drop back to very low power modes - often turning off or power gating the cores. Race-to-halt attempts to reduce the delay in completing a task as much as possible in order to reduce the static power consumption, thereby consuming significantly less power overall. Similarly, \textit{crawl-to-idle} technique aims to execute the workload(s) as slowly as possible so that the battery usage is reduced slowly. Most systems today use a combination of these two philosophies. 

    System level power and thermal management techniques fall into these broad categories: 
	\begin{enumerate}
	    \item \textit{Idle power management} is essentially doing nothing, efficiently. Idle power management is used at all levels of hierarchy - from the smallest part of the circuit or microarchitecture to processor(s), memory controller, hard drives / SSD, network engine, input/output (IO) subsystems and fabrics/interconnects. OS components - typically device drivers that look at scheduler load, next expected interrupts, and other heuristics guide the system to lowest possible idle state through fine grained orchestration among software and hardware.
	    \item \textit{Active power management} is the energy efficient operation of the entire system during active workloads. This is used for all parts of the logic that use a dynamic operating voltage range - microprocessors, GPUs, memory subsystem, etc. Many system components run at fixed voltage, hence this does not apply to them; however, \textit{dynamic frequency scaling (DFS)} can be used if that feature is supported. There are several parts of the OS across kernel, device drivers, firmware that coordinate DVFS based on the current workload, utilization of the hardware, and other heuristics.
	    \item \textit{Duty Cycling} - A duty cycle or power cycle is the fraction of one period in which a signal or system is active. In the context of a computing system, or a specific logic block, duty cycling can be used to cycle components on/off based on several considerations - currently executing workload, runtime counters that indicate usage (or expected usage), etc. 
	    \item \textit{Race-to-idle} - The concept here is to keep the system in its highest operating state (frequency/voltage) in order to complete the workload as fast as possible and then go to sleep or its lowest operating state (frequency/voltage).
	    \item \textit{Crawl-to-halt} - Sometimes, depending on the nature of the workload (or phases of the workload), it may not be ideal to run at the highest operating state. For example, in a battery-powered device (IoT/wearable), in order to conserve battery for as long as possible, several system components are kept at lower operating states to maximize usage of the device. This could mean the workload (or system as a whole) will run slower but battery life is extended as much as possible.
	    \item \textit{Thermal management} is active or passive cooling based on the form factor, workloads, and environmental variables. Usually devices operate until the Thermal Design for Power (TDP) point is breached (or close to being breached), at which point, thermal throttling algorithms kick in and attempt to reduce the overall temperature of the device by taking specific actions to reduce the heat dissipation.
	\end{enumerate}

    \subsubsection{Software Guided Power Management}
    Software and operating systems have evolved over time to provide complex, configurable policies and mechanisms for power and performance control of processors and other subsystems. 
    
    For idle power management, OSes such as Linux and Windows detect idleness of the CPU, interconnects, and peripherals to trigger hardware idle states through architectural interfaces. In Linux, for example, the CPU idle subsystem monitors the CPU and rest of the system for idleness, next expected interrupt and system QOS requirements to trigger low power states for the CPU and entire SOC through the MWAIT x86 instruction. In ARM processors, the equivalent interface is the WFI instruction. 
    
    For active power management, current operating systems provide sophisticated control policies and mechanisms for controlling the CPU, GPU, and some peripherals as well. In Linux, CPUFreq is a standard framework used for CPU Dynamic Voltage and Frequency Scaling (DVFS). Processors have a range of frequencies and corresponding voltages over which they may operate. The CPUFreq framework allows for control of these voltage-frequency pairs according to the load, and user controllable policies, through components called \textit{governors}. There are several different governors based on how the algorithm can be controlled and implemented - \textit{performance governor}, \textit{power-save governor}, \textit{user-mode governor}, etc. The \textit{on-demand governor} is one of the most popular governors \cite{linux/on-demand}, which, as the name indicates, controls the CPU DVFS based on the load. The \textit{interactive governor} is suited for touchscreen-based mobile devices that require optimized burst performance for on-screen usages. The \textit{Intel P-state driver} can operate in two different modes, active or passive. In the active mode, it uses its own internal performance scaling governor algorithm or allows the hardware to do performance scaling by itself, while in the passive mode it responds to requests made by a generic CPUFreq governor implementing a certain performance scaling algorithm. All of these are described in detail in the Linux kernel documentation \cite{linux/dvfs} and the Intel P-state driver is described in more detail in \cite{intel/p-state-driver}.

\section{Architectural Trends and System Level Energy Efficiency}
\label{section:archtrends}
    John Hennessy and David Patterson, in their recent ACM Turing award lecture and publication \cite{henessey-patterson-golden-age-comp-arch} trace the history of computer architecture and touch upon some of the recent trends, including domain-specific architectures (DSA), domain-specific languages (DSL) and open instruction set architectures such as RISC-V (Patterson et al. \cite{risc-v-patterson}). In this section, we elaborate on some of the key observations highlighted in Hennessey and Patterson \cite{henessey-patterson-golden-age-comp-arch}, look at how the underlying architecture of computing systems has transformed in the last couple of decades due to several fundamental laws and limits, and focus on system level energy efficiency. Markov \cite{Markov_2014} discusses some of these trends as well, specifically with regard to \textit{limits on fundamental limits to computation}. We will look at the trends, inflexion points and their respective impact on system level energy efficiency detailed in Table \ref{tab:ee-arch-trends-energy-efficiency}. This list is, by no means exhaustive, however it aims to illustrate the influence of key inflexion points on energy efficiency.

 \begin{table}
   	\caption{Trends in system architecture and energy efficiency}
	\label{tab:ee-arch-trends-energy-efficiency}
    \begin{tabular}{|c|l|}
        \toprule
        \textbf{Architectural Trends} & \textbf{Energy Efficiency Features} \\
        \midrule
        \multicolumn{1}{|m{6cm}}{Moore's Law\cite{moore:1965}, ILP wall (David Wall \cite{davidwall-ilplimits}), Dennard Scaling\cite{dennard74design}} & 
		\multicolumn{1}{|m{6cm}|}{Increased performance via superscalar, VLIW arch, Clock/power gating, processor, cache, memory sleep states, Dynamic Voltage Frequency Scaling (DVFS), power delivery improvements} \\ \hline
		\multicolumn{1}{|m{6cm}}{Multi-cores\cite{intelterascale}, Amdahl's limit (Amdahl \cite{Amdahl67}, Hill et al. \cite{amdahl-multicore})} & 
		\multicolumn{1}{|m{6cm}|}{OS guided / controlled sleep states, fine grained clock/power gating, per-core, per-module DVFS, on-die voltage regulators}\\ \hline
		\multicolumn{1}{|m{6cm}}{Memory wall, Phase Change Memory (PCM) \cite{memory-pcm}, Magneto-resistive RAM (MRAM)\cite{intel-stt-mram}, Spin Transfer Torque RAM (STT-RAM)\cite{sttram-main-mem-onur}, Resistive RAM (ReRAM) \cite{ibm-research-mram-2020}} & 
		\multicolumn{1}{|m{6cm}|}{Memory DVFS (Deng et al. \cite{memscale-bianchini}, David et al. \cite{memdvfs-gorbatov}), system level techniques for memory power management\cite{intel-optane}} \\ \hline
			
		\multicolumn{1}{|m{6cm}}{Domain-specific architectures such as programmable network processors (Li et al. \cite{p4-dsa-nw}), deep learning chips (Jouppi et al. \cite{googletpu}), Intel Mobileye\cite{intel/mobileye/eyeq-ultra}} & 
		\multicolumn{1}{|m{6cm}|}{Chip/IP-level clock/power gating, DVFS}\\ \hline
			
		\multicolumn{1}{|m{6cm}}{Dark silicon challenges (Esmaeilzadeh et al. \cite{tocs12:dark-silicon})} & 
		\multicolumn{1}{|m{6cm}|}{Fine grained power domains and islands}\\ \hline
			
		\multicolumn{1}{|m{6cm}}{High bandwidth interconnects} & 
		\multicolumn{1}{|m{6cm}|}{Standards like CXL\cite{cxl} and PCIe \cite{pci-sig}}\\ \hline
			
		\multicolumn{1}{|m{6cm}}{Non von-Neumann architectures (David Culler \cite{culler1986dataflow}, SpiNNaker\cite{spinnaker}, Thakur et al. \cite{snnsurvey})} & 
		\multicolumn{1}{|m{6cm}|}{Energy-aware dataflow architectures} \\ \hline
		
		\multicolumn{1}{|m{6cm}}{Combining von Neumann and non-von Neumann chips (Nowatzki et al. \cite{mix-vn-non-vn})} & \multicolumn{1}{|m{6cm}|}{Emerging area, mix of different techniques} \\ \hline
		
		\multicolumn{1}{|m{6cm}}{Power delivery miniaturization} & 
		\multicolumn{1}{|m{6cm}|}{On-die/chip voltage regulators, software control, reconfigurable power delivery (Lee \cite{rpdn})}\\ \hline
			
		\multicolumn{1}{|m{6cm}}{Programmable architectures - FPGAs} & 
		\multicolumn{1}{|m{6cm}|}{Energy-aware FPGAs, still in nascent stage}\\ \hline
		
		\multicolumn{1}{|m{6cm}}{Energy Proportional Computing\cite{energy-proportional-computing-google}} & 
		\multicolumn{1}{|m{6cm}|}{Energy-aware data centers, system components}\\ \hline
		
        \multicolumn{1}{|m{6cm}}{Near/sub-threshold voltage designs\cite{subthreshold/design}\cite{subthreshold/techniques}, 3D stacking\cite{tsv-intro}, and chiplets\cite{intel-lakefield}} & 
        \multicolumn{1}{|m{6cm}|}{Ultra low voltage designs, Thermal algorithms} \\ \hline			
        \multicolumn{1}{|m{6cm}}{Thermodynamic computing \cite{thermodynamic-computing}, Landauer Limit \cite{landauer-limit} and Quantum Computing \cite{quantum-computing-survey}} &
        \multicolumn{1}{|m{6cm}|}{Emerging areas, system architectures unclear / evolving} \\
        \bottomrule
    \end{tabular}
    \end{table}

    \subsection{Moore's Law, Dennard Scaling and Instruction-Level Parallelism}
    \label{subsection:mooreslaw-ilp-dennard}
    Moore’s Law \cite{moore:1965} has enabled the doubling of transistors on chips approximately every 18 months through innovations in device, process technology, circuits and microarchitecture, and this has in turn spurred several innovations in system software, applications, thermal management, heat dissipation, advanced packaging and extreme form factors. It is interesting to note that Gordon Moore had himself predicted a slowdown in 2003 as CMOS technology approached fundamental limits (Moore \cite{Moore2003NoEI}). In addition to this, there have been other important laws that have shaped computer systems. One such is Dennard Scaling \cite{dennard74design}. Robert Dennard observed in 1974 that power density stays constant as transistors get smaller. The key idea was that as the dimensions of a device go down, so does power consumption. For example, if a transistor’s linear dimension shrank by a factor of 2, that gives 4 times the number of transistors. If both the current and voltage are also reduced by a factor of 2, the power it used would fall by 4, giving the same power at the same frequency. While this law held, smaller transistors ran faster, used less power, and cost less. During the last decade of the 20th century and the first half of the 21st, computer architects made the best use of Moore’s Law and Dennard scaling to increase resources and performance with sophisticated processor designs and memory hierarchies that exploited instruction level parallelism (ILP). Dennard scaling, however, soon ended because current and voltage could not keep dropping while remaining dependable. Recently, near-threshold and sub-threshold voltage technologies \cite{sub-threshold-startups} are attempting to push these boundaries. 
    
    Instruction Level Parallelism (ILP) can be implemented through several different techniques, and the amount of ILP in programs can be application specific. Scientific computing, graphics applications may exhibit high ILP whereas workloads such as cryptography may not. Micro-architectural techniques that are used to exploit ILP include:
    \begin{enumerate}
        \item \textit{Instruction pipelining}: Here the execution of multiple instructions can be partially overlapped thereby reducing the overall Clocks-per-instruction (CPI). 
        \item \textit{Superscalar execution}, \textit{Very Long Instruction Word (VLIW)}, \textit{Explicitly Parallel Instruction Computing (EPIC)}: In these, multiple execution units are used to execute multiple instructions in parallel. In superscalar designs, multiple instructions can be executed in a clock cycle by dispatching multiple instructions to different execution units on the processor (Palacharla et al. \cite{superscalar/jouppi}). There were several variations of this architecture as well, such as the Ultrascalar (Henry et al. \cite{ultrascalar/yale}) and Multiscalar processors (Sohi et al. \cite{multiscalar/sohi}). In Very Long Instruction Word (VLIW) designs, one VLIW instruction encodes multiple operations with at least one operation for each execution unit. Efficiency of VLIW architectures relies heavily on compilers to correctly schedule operations (Fisher \cite{vliw/fisher}). EPIC architectures evolved from VLIW (Schlansker and Rau \cite{epic-arch/hp}), but retained many concepts of superscalar architectures, and formed the foundation of many generations of Intel processors, including Itanium. While VLIW and EPIC architectures did not gain popularity in mainstream processors, some domain-specific chips have used VLIW architectures. For example, AMD's TeraScale GPU \cite{amd/terascale-gpu} was based on VLIW and more recently, Intel's Movidius \cite{intelmovidiusmyriad2} is a VLIW-based low power inference chip.
        \item \textit{Out-of-order execution}: In this, instructions execute in any order as long as they do not violate data dependencies. This can be implemented on any of the above architectures (pipeline-based on superscalar). 
        \item \textit{Register renaming} is used to avoid unnecessary serialization of program operations when hardware registers are used to store program operands. This technique, originally devised as Tomasulo's algorithm \cite{tomasulo}, is widely used in almost all processor architectures today. 
        \item \textit{Speculative execution}: This allows the execution of instructions before being certain whether the instruction would be executed, and is implemented by using techniques such as control flow speculation, memory dependence prediction, etc. 
        \item \textit{Branch prediction}: This is used to avoid stalling, and is heavily used with speculative execution. 
    \end{enumerate}
    
    While some of these ILP improvement techniques ran out of steam due to various reasons \cite{davidwall-ilplimits}, many of these techniques are still used today in modern processors. 
    Even as ILP limits were worked around through afore mentioned techniques, the industry started to switch from single energy-hogging processors to multiple efficient processors or many cores per chip, ushering in the many/multi-core era. Recent times have also seen hybrid designs that combined low power/low performance and high power/high performance cores, like ARM's BIG.LITTLE architecture \cite{big-little-ee} and the recent Intel Lakefield chip \cite{intel-lakefield}. Hameed et al. \cite{inefficiencies-gp-processors} explore the sources of performance and energy overheads of common workloads on a general purpose CMP system, and look into methods to eliminate these overheads by customizations to CPU cores. The general approach is that as ASICs are significantly more energy efficient than general purpose CMP systems, achieving comparable energy reduction requires algorithm-specific optimizations, such as specialized functional units. Even as Moore's Law slows down, transistor density scaling has continued to be exponential, as illustrated in Figure \ref{fig:42-years-microprocessor-trends}. Etiemble \cite{etiemble2018-45years-processor-evolution} describes evolution of CPUs over the last 45 years. 
	
    \subsection{Multi-core era, Amdahl's law}
    \label{subsection:multi-core-amdahl}
  There were limits to the multi-core era too, as dictated by Amdahl's law \cite{Amdahl67}, which states says that the theoretical speedup from parallelism is limited by the sequential part of the task; so, for example, if \( \frac{1}{8} \)\textit{th} of the task is serial, the maximum speedup is 8 times the original performance, even if the rest is easily parallelizable and we add any number of processors. Hill et al. \cite{amdahl-multicore} elaborate on the impact of this law on multi-core chips. 
  
  	
  	Let speedup be the original execution time divided by an enhanced execution time. Amdahl's law states that if we enhance a fraction \textit{f} of a computation by a speedup S, the overall speedup is: \\
  	$$
 	{\rm Speedup}_{{\rm en}{\rm ha}{\rm nc}{\rm ed}}(f,S)= \frac{1}{(1-f) + \frac{f}{S}} \\
 	$$
 	More specifically, if we are using \textit{n} processor cores: \\
 	$$
 	{\rm Speedup}_{{\rm parallel}}(f,n)= \frac{1}{(1-f)+\frac{f}{n}} \\
 	$$
     
	
\subsection{The Problem of Dark Silicon}
\label{subsection:dark-silicon}
For decades, Dennard scaling permitted more transistors, faster transistors, and energy efficient transistors with each new process node, justifying the costs required to develop each new process node. Dennard scaling’s failure led the industry to race down the multicore path, which for some time permitted performance scaling for parallel and multitasking workloads, permitting the economics of process scaling to hold. The next problem that all chips have had to deal with over the last decade is that of \textit{dark silicon}. Several studies, like Esmaeilzadeh et al. \cite{tocs12:dark-silicon} show that regardless of chip organization, architecture or topology, the runtime software (at OS/firmware/hardware levels) must essentially shut off several parts of the silicon due to fundamental power and thermal limits. This part of the hardware is termed as \textit{dark silicon}. 

Due to dark silicon, even if the increased number of transistors is used to implement additional processor cores, not all available cores can be powered at the same time, to avoid overloading the thermal budget of the chip. Recent designs, especially in power-constrained devices, use dedicated co-processors to run particular tasks in a power-optimized fashion that can be turned off when not in use. These designs rely heavily on aggressive power gating, dynamic voltage and frequency scaling and are organized into fine-grained power and voltage domains. Software and operating system guided energy efficiency is all the more paramount since higher layer of intelligent software should devise strategies for aggressively powering on/off different components of the system based on the usage scenario.


Other techniques are being explored as well to mitigate the effect of dark silicon. Asynchronous circuits is one such technique. While synchronous circuits use a single global control signal, which is active at times even when there is no processing needed in a particular pipeline, asynchronous circuits are only active when workloads are in local pipelines. Techniques like desynchronization are used to convert a synchronous circuit into an asynchronous one. Boundary synchronization is another technique that is used to perform synchronization of signals as they cross clock and voltage domains. Krstic et al. \cite{async-circuits} provide a detailed survey of \textit{Globally Asynchronous, Locally Synchronous (GALS) circuits}. Another method for reducing power consumption in asynchronous circuits is \textit{energy modulated computing} (Yakovlev \cite{energy-modulated-computing}). Here, asynchronous logic uses the power available to it and adjusts the performance to meet that energy level.  
    
\subsection{Memory wall, improved memory technologies}
\label{subsection:mem-wall}
    Dynamic Random Access Memory (DRAM) has been the mainstay of memory systems over the last few decades across almost all computing systems. As applications/workloads evolve, the data set sizes have rapidly grown, along with an increase in the need for rapid analysis of such data. Moving data from memory to the processing unit and back turned out to be a limiting factor for both performance and power consumption, especially for workloads such as deep learning that involve repetitive operations on large data sets. This limiting factor is termed the \textit{von Neumann bottleneck}, or \textit{memory wall}, which is essentially the bottleneck imposed by the bandwidth of the channel between the CPU/GPU or accelerator and the memory subsystem. While GPUs were a good fit for the computational elements of deep learning algorithms, the limitations from the memory wall proved to be the next obstacle to overcome. For these real-time big data workloads, DRAM was not big enough and traditional storage was not fast enough. 
    

    DRAM scaling faces significant challenges; Mandelman et al. \cite{ibm-dram-scaling} describe how the scaling techniques used in earlier generations are encountering limitations that require major innovations. Mutlu \cite{onur-multu-mem-scaling-systems} describes in detail the demands and challenges faced by the memory system, and examines some recent research and industrial trends to overcome these challenges, primarily around new DRAM architectures, better integration of DRAM with the rest of the system, designing new memory systems that employ emerging memory technologies and providing predictable performance and QoS as workloads become more data-movement intensive. 
    
    Improvements in memory technology over the last two decades has focused on newer memory technologies, improving energy efficiency of the memory systems, and more recently, embedding logic closer to, or along with, memory. Each of these are now briefly described. 
    
    \subsubsection{Newer Memory Technologies}
    \label{subsubsection:new-memory}
    Phase Change Memory (PCM) (Lee et al. \cite{memory-pcm}) is an alternative to DRAM and provides a nonvolatile storage mechanism that scales well. Raoux et al. \cite{ibm-pcm-report} discuss the critical aspects that may affect the scaling of PCM-based RAM, including material properties, power consumption during programming and read operations, thermal cross-talk between memory cells, and failure mechanisms. Wong et al. \cite{pcm-report} survey the electrical and thermal properties of phase change materials with a focus on the scalability of the materials and their impact on device design. The authors also provide an in-depth review of innovations in device structure, memory cells, strategies for achieving 3D high-density memory arrays. Scalability implies lower main memory energy and greater write endurance. In the original PCM architecture, during writes, an access transistor injects current into the storage material and thermally induces phase change, which is detected as a programmed resistance during reads. Since PCM relies on analog current and thermal effects, it does not require control over discrete electrons. As technologies scale and heating contact areas shrink, programming current is expected to scale linearly. Starting with a 32nm device prototype, this has now led to generations of products in the industry including Numonyx, Western Digital, Samsung and Intel/Micron's 3D Xpoint memory. Intel, for example, has developed two different ranges of mass volume products based on PCM \cite{intel-optane} - standards-based PCIs Optane Solid State Drives (SSD)s that are broadly compatible with a wide range of systems, and as an on-board memory caching/acceleration device. 
    
    While PCM-based technologies like Intel Optane have started seeing deployments in datacenters, magneto-resistive RAM (MRAM) has shown promise for low power IoT devices. Data in MRAM is not stored as electric charge or current flows, but by magnetic storage elements formed from two ferromagnetic plates, each of which can hold a magnetization, separated by a thin insulating layer. One of the two plates is a permanent magnet set to a particular polarity; the other plate's magnetization can be changed to match that of an external field to store memory. This configuration is known as a \textit{magnetic tunnel junction} and is the simplest structure for an MRAM bit. A memory device is built from a grid of such "cells". MRAMs are non-volatile like PCM, very fast with read/write latencies of around 35 nanoseconds, reliable at different temperatures, consume very low power and can be manufactured at leading process nodes - IBM recently announced a 14nm MRAM node \cite{ibm-research-mram-2020}. However, due to its lower density, MRAMs is expected to be more applicable to smaller IoT devices.
    
    Non-volatile Spin Transfer Torque Random Access Memory (STT-RAM) (Chen et al. \cite{sttram-ref}) combines the capacity and cost benefits of DRAM, the fast read and write performance of SRAM and the non-volatility of Flash memory with improved endurance. STT-MRAM is a variation of MRAMs and it works by controlling electron spin with a polarizing current, requiring less switching energy than MRAMs, thereby bringing power consumption down. It has near-zero leakage power, lower active power consumption scalability and simpler manufacturing beyond 45nm. Kültürsay et al. \cite{sttram-main-mem-onur} showed that an optimized, equal capacity STTRAM main memory can provide performance comparable to DRAM main memory, with an average 60\% reduction in main memory energy. STT-RAM can be added to memory controllers to improve caching performance and power-loss protection, and can be scaled with technology nodes - Intel recently announced their production-ready STT-RAM array integrated with 22nm process \cite{intel-stt-mram}.
    
    Resistive RAM (ReRAM) (Bhattacharjee et al. \cite{rram-ref}) is another type of non-volatile memory that works by changing the resistance across a dielectric solid-state material, also known as \textit{memristor}. It consumes low power, exhibits high density, and a performance profile that makes it amenable to structure it in between DRAM and flash-based storage. While MRAM’s characteristics make it more suitable for IoT devices, ReRAM bridged the gap between server memory and SSDs thereby making it a candidate for datacenters. Another interesting facet of ReRAMs (and memristors in general) is that they mimic the human brain's biological computation at the neuron and synaptic level. Mehonic et al. \cite{bipin2020memristors} describe how memristor technology has the potential to scale computation beyond traditional von-Neumann computing models and can help with energy-efficient deep learning accelerators and spiking neural network based architectures. 
    Wong et al. \cite{memory/metal-oxide-rram} describe the physical mechanism, material properties and electrical characteristics behind binary metal-oxide resistive RAM. Due to the improvements in endurance, retention, multi-bit operation and scalability, large-scale RRAM arrays are now possible. 
    
    \subsubsection{Energy Efficiency Techniques for Memory Systems}
    \label{subsubsection:ee-memory}
    Several techniques to optimize DRAM-based systems have been explored and implemented in research and commercial systems, a few of which are highlighted here. Lee et al. \cite{memory/power-perf-dram} describe the power and performance relationship of modern DRAM devices. In \cite{memory/ee-dram} and \cite{memory/dram-half-page}, Ha et al. provide an exhaustive analysis of state-of-the-art DRAMs, LPDDR4, HBM and describe the design and implementation of several energy reduction techniques by optimizing accesses (Half page DRAM technique reduces energy by 38\%) and refresh cycles (Charge Recycling Refresh technique conserves 32\% energy and Smart Refresh improves this even further). Similarly, Liu et al. \cite{memory/raidr} propose RAIDR (Retention-Aware Intelligent DRAM Refresh), a mechanism that can identify and skip unnecessary refreshes using knowledge of cell retention times. This is done by grouping DRAM rows into retention time bins and applying a different refresh rate to each bin, thereby reducing the refresh cycles of less frequently used bins/cells. This technique achieves an impressive 74\% refresh reduction leading to a DRAM power reduction of 16\%. Chang et al. \cite{dram-voltage-reductions} explore reducing the DRAM supply voltage more aggressively to reduce energy consumption by studying about 125 real LPDDR3 DRAM chips. They find that while reducing supply voltage introduces bit errors, they can be avoided by increasing the latency of key DRAM operations such as activation, restoration and precharge. They also propose a technique called Voltron, which uses a performance model to determine how much the supply voltage can be dropped without errors. These improvements outperform previous DRAM DVFS algorithms for memory intensive workloads. 
    
    Going beyond the DRAM subsystem itself, several approaches to using DVFS have been researched and implemented, both for the memory subsystem itself, as well as coordinated DVFS across CPU, memory and other subsystems. The advent of Memory DVFS, which is the ability to dynamically scale the voltage and frequency of the memory subsystem, independent of CPU DVFS, allowed for optimizing the system as a whole from an energy efficiency perspective. There have been several approaches to this. One of the earliest approaches was to adjust CPU DVFS based on memory accesses, so that the memory subsystem could enter idle low power states if the CPU was busy executing computations and there were no pending memory operations. For example, Liang et al. \cite{ma-dvfs} demonstrated how performance monitoring counters could be used to alter CPU DVFS and help lower system energy consumption in an embedded device. Howard et al. \cite{memdvfs-gorbatov} was one of the earliest works in memory DVFS that demonstrated a simple control algorithm that adjusts memory voltage and frequency based on memory bandwidth utilization, and was implemented on a real system. Deng et al. \cite{memscale-bianchini} describe MemScale, a technique that leverages dynamic profiling, performance and power modeling, DVFS of the memory controller, and DFS of the memory channels and DRAM devices, all done independent of CPU DVFS. MemScale is guided by an operating system policy that determines the DVFS/DFS mode of the memory subsystem based on the current need for memory bandwidth, the potential energy savings, and the performance degradation that applications are willing to withstand. Bianchini et al. \cite{coscale-memdvfs-bianchini} describe CoScale, for coordinating memory and CPU DVFS in server systems. CoScale relies on execution profiling of each core through performance counters, and models of core and memory performance and power consumption. It uses fixed-size epochs (matching an OS time quantum). In each epoch, there is a system profiling phase followed by the selection of core and memory subsystem frequencies that  minimize total system energy while maintaining performance within the target bound. The advent of GPUs and memory bandwidth hungry workloads extended this concept to coordinated CPU, GPU and memory DVFS using performance and power monitoring counters. Chau et al. \cite{ee-cpu-gpu-dvfs} describe a scheduling algorithm that optimizes the CPU and GPU DVFS states based on currently running workloads and their predicted runtime. Most recent systems from Intel, AMD, NVIDIA, etc. support independent DVFS for CPU, GPU, memory, along with techniques such as dynamic memory throttling (inserting idle cycles between reads and writes). Mittal and Vetter \cite{cpu-gpu-dvfs-vetter-mittal} present a survey of these CPU-GPU coordination techniques.
    
    \subsubsection{Processing In Memory (PIM)}
    \label{subsubsection:processing-in-memory}
    With recent advances in existing memory systems, and the advent of newer memory techniques, integration of memory and logic, an old idea, has re-emerged. Mutlu et al. \cite{mutlu-pim-primer} describe this in exhaustive detail and we use the same terminology here. Broadly this is called \textit{processing-in-memory} and it involves placing computation mechanisms in or near where the data is stored (memory chips, or the logic layer, or the memory controllers, etc.), so that data movement is reduced or eliminated. \textit{Processing-In-Memory (PIM)} is also called \textit{Near-Data-Processing}. PIM involves the following categories of techniques:
    \begin{enumerate}
        \item \textbf{Processing using memory} - In this category, the idea is to improve overall energy efficiency and performance.  Boroumand et al. \cite{google-workloads-pim} analyze the energy and performance impact of data movement for several widely-used Google consumer workloads such as Chrome browser, TensorFlow Mobile (Google's ML framework) and video playback/capture. The authors observe that data movement accounts for almost 63\% of the total energy consumed. Further, as most of the data movement is generated from simple functions/primitives (such as memcopy, memset), implementing these primitives in PIM hardware reduces the system energy by almost 55\% with a corresponding 54\% increase in system performance. Shuangchen et al. \cite{drisa-dram} propose DRISA, a DRAM-based Reconfigurable In-Situ Accelerator architecture that uses DRAM memory arrays that can be reconfigured to compute various Boolean logic functions. They also optimize for high performance by simultaneously activating multiple rows and sub-arrays, thereby providing massive parallelism and unblocking internal data movement bottlenecks, leading to improved performance and energy consumption at the system level as compared to ASICs and GPUs. Similarly, Seshadri et al. \cite{memory/rowclone} and \cite{memory/ambit-dram}  propose two different techniques that can be used in existing DRAM systems with minium modifications. Seshadri et al. \cite{memory/rowclone} propose RowClone, a mechanism to perform bulk copy and initialization completely within DRAM by optimizing copy operations between rows and also by using the shared internal bus of a DRAM chip to copy data between two banks. These techniques yield a 11X latency reduction and 75X energy reduction for typical copy operations. Seshadri et al. \cite{memory/ambit-dram} propose Ambit, an easy to implement architecture for existing DRAM systems to optimize bulk bitwise operations, which are the major component of database, websearch and neural network workloads. Ambit is an \textit{acclerator-in-memory}; with minimum changes to the DRAM sense amplifiers, existing DRAM can perform bulk bitwise operations. With these modifications, Ambit shows a 32x performance improvement and 35X energy reduction. Integration with HMC improves the bulk bitwise operations by 10X as well. For accelerating deep neural network workloads (CNNs and DNNs), Shafiee et al. \cite{memory/isaac-dram} propose ISAAC, an In-Situ Analog Arithmetic in Crossbars. The key idea in this technique is to use the memristor crossbar array not only to store input weights, but also to perform dot-operations in an analog manner. They design a pipelined architecture, with some crossbars dedicated to each neural network layer, and eDRAM buffers that aggregate data between pipeline stages. ISAAC demonstrates a 15X improvement in throughput and a 5X reduction in energy consumption. Exploring similar ideas for non volatile memories, Shuangchen et al. \cite{memory/pinatubo-pim} propose Pinatubo, a PIM architecture for bulk bitwise operations in non volatile memories. In this technique, they redesign the read circuitry so that it can compute bitwise logic of two or more memory rows efficiently and can perform one-step multi-row operations. Pinatubo demonstrates a 500X speedup and 28000X energy savings as compared to conventional systems. Prezioso et al. \cite{memory/memristor-crossbar} discuss similar techniques for crossbar architectures for neuromorphic metal-oxide memristor circuits. They make the observation that the extreme complexity of the human cerebral cortex makes the hardware implementation of neuromorphic networks with a comparable number of devices exceptionally challenging. CrossNets based on hybrid CMOS/memristor circuits, where CMOS stack is augmented with crossbar layers, seem promising, even though each crosspoint requires an additional transistor. Ping et al. \cite{prime-reram} propose using ReRAM for main memory. Given ReRAM's crossbar array structure, ReRAM-based memory can perform matrix operations efficiently and is therefore interesting for neural network workloads. The authors propose a novel PIM architecture called PRIME, where a portion of ReRAM crossbar arrays can be configured as accelerators for NN applications or as normal memory for a larger memory space, along with a hardware/software interface to optimize neural workloads. Their results demonstrate a 2360X improvement in performance with a 895X improvement in energy consumption, when compared with a state-of-the-art neural processing unit design. Ambroglio et al. \cite{ambrogio2018equivalent} describe an analogue non-volatile memory implementation for accelerated neural network training. Analogue non-volatile memory can accelerate the neural-network training algorithm known as backpropagation by performing parallelized multiply–accumulate operations in the analogue domain at the location of the weight data. The authors demonstrate mixed hardware–software neural-network implementations that combine long-term storage in phase-change memory, near-linear updates of volatile capacitors and weight-data transfer with ‘polarity inversion’ to cancel out inherent device-to-device variations. The techniques demonstrate a 2X improvement of energy efficiency and performance as compared to traditional GPUs. 
        
        \item \textbf{Processing near memory} - The idea behind this technique is to take advantage of computation capability in conventional memory controllers or the logic layer(s) of 3D-stacked memory technologies. This can be done with current memory and packaging technologies such as 3D stacking. 3D stacking technology is becoming an instrument for scaling system performance and densities because of increased inter-layer bandwidth, reduced inter-layer latency and ability to integrate dies from different process technologies as a means of customization (Knickerbocker et al. \cite{knickerbocker2008three}). 3D-stacking technology enables the integration of DRAM and logic offering high bandwidth and reduced energy consumption. Architectures such as High Bandwidth Memory (HBM) (Lee et al. \cite{lee20151}) and Hybrid Memory Cube (HMC) (Hadidi et al. \cite{hmc1}) are some of the recent near-memory compute systems built with 2.5D/3D stacking. HMC is a 3D-stacked DRAM device and comprises of several DRAM dies with a logic layer connected vertically with \textit{Through-Silicon-Vias} (TSVs) (Hadidi et al. \cite{hmc1}, Pawlowski et al. \cite{hmc2}). Energy efficiency is achieved by offloading tasks onto bandwidth-rich processing units embedded in the logic layer of the 3D-stacked memory. It has a memory hierarchy that enables large number of simultaneous memory accesses. Each DRAM layer in HMC is divided into a number of equal partitions. Vertically aligned partitions of all layers form a \textit{vault}. All vaults are functionally and operationally independent from each other and are further divided into banks. Such a hierarchical architecture allows high \textit{memory level parallelism} (MLP) in the hardware which can be exploited well by applications that require it. Also, HMC layers are connected through TSV links which are equivalent to shortened interconnection paths with reduced connectivity impedance allowing higher data movement rates with lower energy-per-bit. Additionally, there are controllers placed in the memory system itself giving HMC the freedom to interact with the memory array more efficiently based on data location (device, vault, bank, row and column) and memory-device timing parameters. Thus, HMC is a promising solution for achieving high energy efficiency for memory-intensive applications. Junwhan et al. \cite{memory/pim/tesseract} propose Tesseract, a programmable PIM accelerator for large scale graph processing using 3D integration. Tesseract is composed of new hardware, an efficient method of communication between different memory partitions and a programming interface to exploit the unique hardware design. They achieve 10X performance improvement and 87X energy reduction over conventional systems across state-of-the art graph processing workloads with large real-world graphs. 
        
    \end{enumerate}
    Taking a completely new approach, Mohamed et al. \cite{memory/N3XT}, describe N3XT, a completely re-architected system using new logic and memory technologies, 3D integration with fine-grained connectivity and new architectures for computation in memory. N3XT uses 1D carbon nanotubes, ReRAMs, STT-MRAM, 3D integration of computing and memory, embedded cooling techniques and new microarchitectures and system runtimes. The authors demonstrate the effectiveness of N3XT by using system-level \textit{energy-delay product} (EDP) - the product of total energy consumption and total execution time. Experimental prototypes of the N3XT technologies demonstrate 10-100X EDP benefits.

    \subsection{Domain Specific Architectures and the limits of chip specialization}
    \label{subsection:dsa-chip-specialization}
    \textit{Domain Specific Accelerators} (DSA) are architectures that are tailored to a specific problem domain and offer significant performance and efficiency gains for that domain. Some examples are GPUs, neural network processors for deep learning and programmable network processors for high speed packet forwarding in software-defined networks (SDNs). 
    
    DSAs for high speed packet processing accelerators have been implemented over the decades starting with ASICs/DSPs to FGPAs and dedicated programmable network processors. In these core internet routers and switches, data plane algorithms must be implemented in hardware in order to do packet processing at line rate of 100s of Gigabits/sec, and they must also be programmable. Several generations of such programmable networking devices form the internet backbone today. Li et al. \cite{p4-dsa-nw} discuss P4GPU for high speed packet processing. The P4 language is an emerging domain-specific language for describing the data plane processing at a network device. P4 has been mapped to a wide range of forwarding devices including NPUs, programmable Network Interface Chips (NICs) and FPGAs. In Sivaraman et al. \cite{packet-transactions-nick}, the authors show how to program data-plane algorithms in a high-level language and compile those programs into low-level microcode that can run on emerging programmable line-rate switching chips using the notion of \textit{packet transactions}, an atomic packet-processing sequence of code. 
    
    Domain specific accelerators for camera/imaging, deep learning, amongst others, have been implemented in several industrial devices in the last 15 years. For example, Qualcomm's Snapdragon SOC contains a Hexagon cores \cite{qchexagon} for AI processing in camera, voice, VR and gaming applications. PowerVR's Neural Net Accelerator (NNA) \cite{powervrnna} is used in several phones/devices. Similarly, Apple's Neural Engine is an AI accelerator core within the Apple recent Bionic SoC \cite{applea12}. Google has built the Tensor Processing Unit (TPU) (Jouppi et al. \cite{googletpu}) that is an ASIC optimized for machine learning and is specifically designed for its TensorFlow framework, which is extensively used for CNNs. Similarly, Intel's Myriad 2 \cite{intelmovidiusmyriad2} is a many-core VLIW AI accelerator complemented with video fixed function units and is reported to be capable of operating in the sub-1W range and delivering 300 GOPS or just over 1 TOPS per watt \cite{intelmovidiusmyriad2powerperf}. Intel's Mobileye's EyeQ \cite{intel/mobileye/eyeq-ultra} is a processor specialized for vision processing for self-driving cars.
	
	The trend of domain specific architectures continues especially in the areas of AI/ML, edge computing for vision/recognition, low power audio processing, and several others. However, much of the benefits of chip specialization stem from optimizing a computational problem within a given chip’s transistor budget. As detailed in Fuchs and Wentzlaff \cite{Fuchs2019TheAcceleratorWall}, for 5nm CMOS chips, the number of transistors can reach 100 billion; however not all of them can be utilized due to the challenge of dark silicon. Chips will be severely limited by thermal budgets. This will also cause stagnation of the number of useful transistors available on a chip, thereby limiting the accelerator design optimization space, leading to diminishing specialization returns, ultimately hitting an \textit{accelerator wall} in the near future. 
	
	\subsection{SOC Integration, evolution of software power management}
	\label{subsection:soc-integ-sw-pm}
	 Computing systems have transformed from predominantly CPU-based systems to more complex system-on-a-chip (SOC) based ones with highly integrated single/multi-core CPUs, newer memory technologies/components, domain-specific accelerators for graphics, imaging, deep learning, high speed interconnects/ peripherals and multi-comms for connectivity. The more recent Compute Express Link (CXL) \cite{cxl} is an industry standard to integrating accelerators, memory and compute elements. Similarly, PCI \cite{pci-sig} have emerged as standards for high bandwidth, low power interconnects between CPU cores, memory and accelerators. As systems have become more capable in terms of their performance and capabilities, their energy consumption and heat production has also grown rapidly. The explosion of highly powerful and complex SOCs across all kinds of computing systems have surpassed the rate of evolution of software thereby presenting unique challenges to meet the power and thermal limits. From a systems perspective, such platforms present wide ranging issues on SOC integration, power closure/verification, hardware/software power management and fine-grained thermal management strategies. \textit{This is perhaps a unique phase in the semiconductor industry which has always prided on a specific cadence of hardware growth and the assumption that software will always be ready to meet the requirements of the hardware}. In order to meet the needs of complex SOCs, operating system and software-guided power management infrastructures, frameworks, and algorithms have evolved different hardware/software techniques. Embedded real time operating systems and open source operating systems such as Linux have developed several software techniques and frameworks to perform aggressive system level power, performance and thermal management such as tickless scheduling (Siddha et al. \cite{linux/tickless}), DVFS frameworks \cite{linux/dvfs}, idle power management (Pallipadi \cite{pallipadi07cpuidle}), active/runtime power management \cite{linux/runtimepm}, and various energy efficient system standby states. Windows has also standardized Connected Standby, Modern standby \cite{ms-modern-standby} and several more energy efficiency strategies and algorithms to manage idle and active workloads. Both Linux and Windows kernel device drivers also implement aggressive energy management techniques at the system level through workload aware PCI link power management (to put internal PCI links in low power state dynamically), network/communications power management, only to name a few. Thus, software guided and software controlled energy efficiency have gained significant importance for complex SOCs and systems. 
    
    \subsection{Advent of non-von Neumann architectures}
    \label{subsection:non-von-nm}
    Traditional architectures have largely followed the von Neumann computing model. One of the major deviations from von Neumann architectures, \textit{dataflow machines} were proposed a couple of decades ago (Culler \cite{culler1986dataflow}) and Veen \cite{veen1986dataflow}. However, they were severely limited by the availability of data movement infrastructures, effective software parallelism and functional units in hardware (Gurd et al.  \cite{gurd1987performance}). Thus, dataflow machines did not see commercial deployment for general purpose computing. However, dataflow architectures have been used significantly in implementing specialized hardware for digital signal processing (DSP), graphics processing, imaging/video engines, etc.
    
    More recently, dataflow or near-dataflow architectures have been applied to AI/ML workloads. Deep learning workloads are largely free of control flow and are instead steered by availability of data for executing a predetermined set of operations. Embodying this algorithmic characteristic, dataflow based systems are being developed which are completely controlled by data flow and not by control. The algorithmic parallelism that such workloads exhibit makes them perfect candidates for dataflow modeling which has the potential of reducing energy consumption by orders of magnitude as compared to their execution on control flow based systems. Most architectures for deep learning acceleration work towards optimizing the data size or the number of operations to be performed which may hold relevance for better performance but do not necessarily translate into energy efficiency. As discussed in Yang et al. \cite{bg6}, there are two reasons to this, data movement and not the computation requires more energy and that the flow of data along with the levels in memory hierarchy have a major impact on energy efficiency. 
    
    Chen et al. \cite{eyeriss-dataflow} propose Eyeriss, an optimized algorithmic dataflow for CNNs by exploiting local data reuse and optimization of intermediate data movement. Tetris (Gao et al. \cite{bg3}) uses the dataflow model of Chen et al.  \cite{eyeriss-dataflow} along with scheduling and partitioning in software to implement CNN acceleration in HMC. In Farabet et al. \cite{farabet2011neuflow}, the authors present Neuflow, a compiler that transforms high level dataflow graphs into machine code representations. Li et al. \cite{li2018smartshuttle} present SmartShuttle, a framework that adaptively switches among different data reuse schemes and the corresponding tiling factor settings to dynamically match different convolutional layers. Its adaptive layer partitioning and scheduling scheme can be added on existing state-of-the-art accelerators to enhance performance of each layer in the network. The industry has also seen some innovative products in this space. Wave Computing (Chris Nichol \cite{nicol2017coarse}), Chaudhuri et al. \cite{chaudhuri2017sat}) present an implementation of a dataflow architecture as an alternative to train and process DNNs for AI especially when models require a high degree of scaling across multiple processing nodes. Instead of building fast parallel processors to act as an offload math acceleration engine for CPUs, Wave Computing's dataflow machine directly processes the flow of data of the DNN itself.
    
    Spiking Neural networks (SNN) are another form of brain-inspired networks that takes a step closer in mimicking the working of the brain. The pulse width and timing relationship between signals adds to the value of the data being computed and SNNs precisely work with these kind of network parameters. Thus, implementations of such neuromorphic loads fall in the larger circle of non-von Neumann computing that are largely asynchronous event-driven systems.

    Some of the implementations of the SNN computing acceleration include IBM TrueNorth (Merolla et al. \cite{akopyan2015truenorth}), SpiNNaker from the University of Manchester (Plana et al. \cite{furber2014spinnnaker}), Intel's Loihi (Davies et al. \cite{intelloihi}) and many more. IBM TrueNorth (Merolla et al. \cite{akopyan2015truenorth}) is a many-core processor network on a chip design, with 4096 cores, each one having 256 programmable simulated neurons for a total of just over a million neurons. In turn, each neuron has 256 programmable ``synapses" that convey the signals between them. Since memory, computation, and communication are handled in each of the 4096 neurosynaptic cores, TrueNorth circumvents the von Neumann architecture bottleneck and is very energy-efficient, consuming 70 milliwatts with a power density that is 1/10,000th of conventional microprocessors. SpiNNaker   (Plana et al. \cite{furber2014spinnnaker}) is a digital neuromorphic neural array designed for scalability and energy efficiency by incorporating brain-inspired communication methods. It can be used for simulating large neural networks and performing event-based processing for other applications. Each node is made of 18 ARM968 processor cores, each with 32 kilobytes of local instruction memory, 64 kilobytes of local data memory, packet router, and supporting circuitry. A single node consists of 16,000 digital neurons consuming 1W of power per node. Ganguly et al. \cite{Ganguly2019TowardsEE} discuss these and other non-von Neumann architectures in more detail with respect to their energy efficiency. 
 
    \subsection{Architectures mixing von Neumann and non-von Neumann chips}
    \label{subsection:vn-non-vn}
\begin{figure}[ht]
		\centering
		\includegraphics[width=\linewidth]{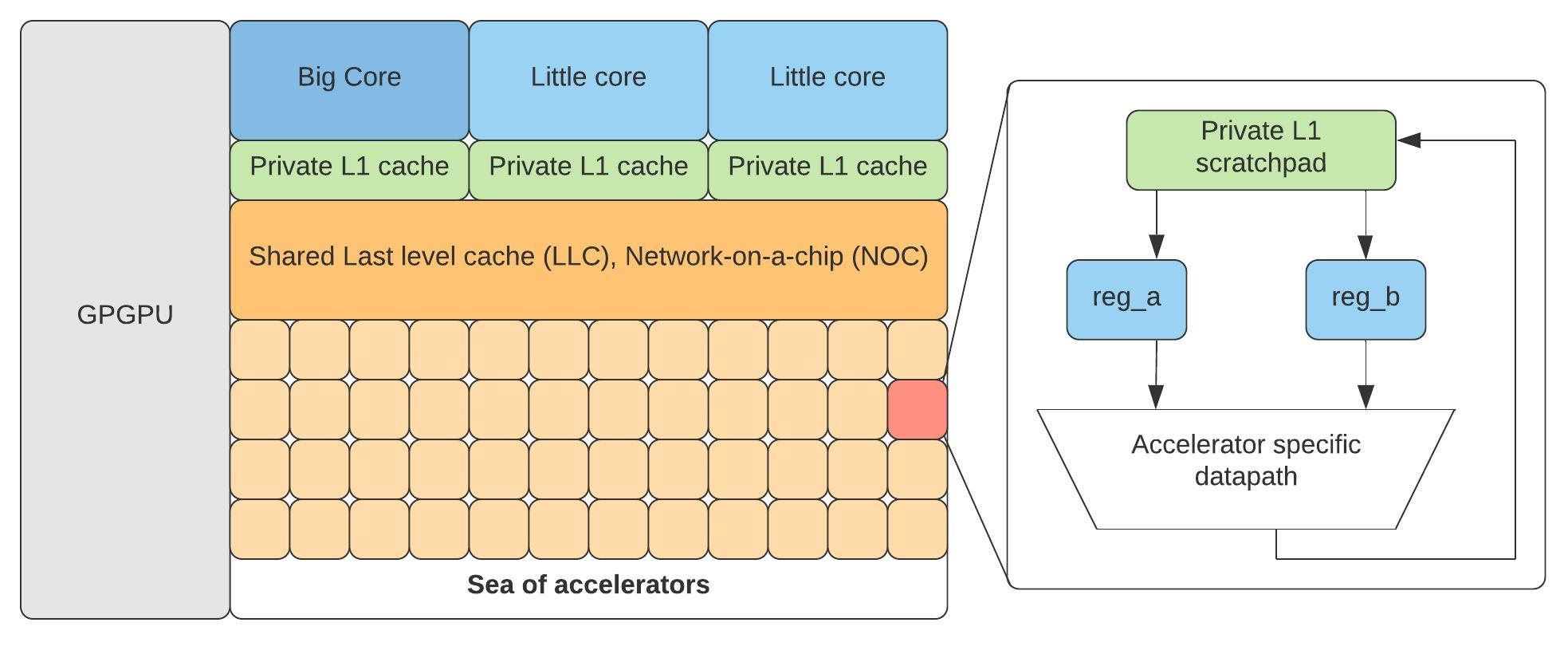}
		\caption{Future heterogeneous architectures \cite{aladdin-accel-sim}} 
		\label{fig:future-hetero-architectures}
	\end{figure}

    With non-von Neumann computing models gaining traction, mixing von Neumann and non-von Neumann architectures/computational models is also being explored. Nowatzki et al. \cite{mix-vn-non-vn} discussed that if both out-of-order and explicit-dataflow were available in one processor, the system can benefit from dynamically switching during certain phases of an application's lifetime. They present analysis that reveals that an ideal explicit-dataflow engine could be profitable for more than half of instructions, providing significant performance and energy improvements. More recently, Intel's Configurable Spatial Accelerator (CSA) \cite{intelcsa} is an effort to mix von Neumann and non-von Neumann processors. The core idea is that there is basic control of data flow (the traditional von Neumann model) but there is also a configurable way to program dataflow parts of the computations. The system takes the dataflow graph of a program (created by compilers) before it is translated down to a specific processor’s instruction set, data storage, and lays down that data flow directly on a massively parallel series of compute elements and interconnects between them. The architecture presents very dense compute and memory, and also very high energy efficiency because only the elements needed for a particular dataflow are activated as a program runs, with all other parts of the chip going idle. The configurable part is that the system will have many different CSA configurations tuned to the dataflows of specific applications (single precision, double precision floating point, mixture of floating point and integer). This is intended to be the first exascale machine deployed in the USA by 2021. It is largely expected that future architectures will be a mix of CPUs, GPUs and domain-specific accelerators, each optimized for a specific function, as shown in Figure \ref{fig:future-hetero-architectures}. Such diverse architectures also make it imperative for the industry and academia to come together and define uniform interfaces across hardware and software to model, estimate, measure and analyze power, performance and energy consumption across layers. Efforts such as the IEEE Rebooting Computing initiative \cite{ieee-rebooting-computing} could be extended to consider this aspect as well in addition to its existing charter.
	
    \subsection{Power delivery miniaturization, reconfigurable power delivery networks}
    \label{subsection:power-delivery}
    From a power delivery perspective, voltage regulators have shrunk and SOCs today have on-die voltage delivery that can deliver fine grained power to different parts of the chip, all of which are controlled through hardware and firmware (and in some architectures, to the OS level as well). SOCs are organized into "power domains" or "voltage islands", which allow for several individual areas of the chip to be powered on/off or run at different clock frequencies/voltage. Haj-Yahya et al. \cite{isqed-2019-powerdelivery} review on-chip, integrated voltage regulator (IVRs) and presents a thorough and quantitative evaluation of different power delivery networks for modern microprocessors. Miniaturization of power delivery has led to another important area - reconfigurable power delivery networks (Lee \cite{rpdn}). This comprises of a network of voltage/frequency converters, a switch network and a controller that can dynamically route power to different areas of the chip to realize fine-grained (zone-specific) voltage/frequency scaling. This is an emerging area across circuit, architecture, and system-level approaches to optimize power delivery to parts of a chip or the entire system based on the current workload(s).
    
    \subsection{Programmable architectures}
    \label{subsection:programmable-archs}
    Field Programmable Gate Arrays (FPGAs) were once applicable to very specific domains and industries. This has changed in the last few years with FPGAs now being a critical component of data center and cloud systems, as well as edge computing systems (Ovtcharov et al. \cite{ovtcharov2015accelerating}). FPGAs are highly programmable in nature as they contain an array of programmable logic blocks, and a hierarchy of "reconfigurable interconnects". The blocks can be "wired together", like many logic gates that can be inter-wired in different configurations, thus making them ideal candidates for \textit{reconfigurable computing systems} that can run highly diverse workloads. However, energy efficiency of such systems is still in its infancy with no easy or standard ways of hardware/software power management across traditional compute and FPGA subsystems.
    
    \subsection{Energy Proportional Computing}
  \label{subsection:energy-proportional-computing}
    In 2007, the concept of \textit{energy proportional computing} was first proposed by Google engineers Luiz André Barroso and Urs Hölzle \cite{energy-proportional-computing-google}. Energy proportionality is a measure of the relationship between power consumed in a computer system, and the rate at which useful work is done (its utilization, which is one measure of performance). If the overall power consumption is proportional to the computer's utilization, then the machine is said to be energy proportional. Up until recently, computers were far from being energy proportional for three primary reasons. The first is high static power, which means that the computer consumes significant energy even when it is idle. High static power is common in servers owing to their design, architecture, and manufacturing optimizations that favor high performance instead of low power. The second reason is that the various hardware operating states for power management can be difficult to use effectively due to complex latency/energy tradeoffs. This is because deeper low power states tend to have larger transition latency and energy costs than lighter low power states. For workloads that have frequent and intermittent bursts of activity, such as cloud microservices, systems do not use deep lower power states due to significant latency penalties, which may be unacceptable for the application(s). The third reason is that beyond the CPU(s), very few system components are designed with fine grained energy efficiency in mind. The fact that the nature of the data center has changed significantly from being compute bound to being more heterogeneous has now exacerbated the problem and energy proportionality of all components will be an important area of research. 
    
    
    \subsubsection{Data center energy efficiency}
    \label{subsub:datacenter-power}
    One of the biggest challenges for large server farms and data center operators is the increasing cost of power and cooling. Over the past decade, the cost of power and cooling has increased tremendously, and these costs are expected to continue to rise. As reported in 2015, (Hamilton  \cite{aws/james-hamilton-data-center-cost}), power distribution and cooling accounts for 18\% of costs in data centers. The Green Grid consortium \cite{greengrid} defines Power Usage Effectiveness (PUE) a metric used to capture the efficiency of a data center’s cooling and power delivery mechanisms. PUE is defined as the ratio of total amount of energy used by a computer data center facility to the energy delivered to computing equipment. 
    
  	$$
 	{\rm {PUE}={\frac{Total\_Power\_Consumption}{IT\_Power\_Consumption}}} \\
  	$$
    
    An ideal PUE is 1.0. Any energy consumption that goes towards a non computing device in a data center falls in the category of facility energy consumption, or IT power consumption. PUE has become the most commonly used metric for reporting energy efficiency of data centers, with many public cloud vendors like Google, Microsoft and Facebook reporting PUE regularly. However, one problem with this metric is that PUE does not account for the climate in the region the data center is built in. In particular, it does not account for different normal temperatures outside the data center. So, a data center running in a tropical region may have a higher PUE than one running in Alaska, but it may actually be running more efficiently. 
    
    PUE was published in 2016 as a global standard under ISO/IEC \cite{pue-ISO-standard} as well as a European standard \cite{pue-EU-standard}. 
    
    Recent research has looked at the impact of the recent explosion in the range of cloud workloads in data centers. In Gan and Delimitrou  \cite{delimitrou-arch-implications-uservices}, the authors investigate the architectural implications of microservices in the cloud, specifically system bottlenecks and implications to server design. Gan et al. \cite{deathstar-bench} present an open source benchmark for microservices, DeathStarBench, that can measure hardware-software implications for data center systems. In Ayers et al. \cite{asmDB-fe-stalls}, the authors present asmDB, which looks at the source of front end stalls (cache misses, instruction cache misses, etc.) in large warehouse-scale computers, and present some optimizations that can help mitigate such system bottlenecks. Mirhosseini et al. \cite{akshitha-killer-microseconds} explore \textit{killer microseconds} - microsecond-scale "holes" in CPU schedules caused by I/O stalls or idle periods between requests in high throughput microservices that are typical in data centers. They then propose enhancements to server architectures to help mitigate such effects. At a system level, Ilager et al. in \cite{ilager2020thermal} explore using ML techniques for thermal prediction for energy efficient management of cloud computing systems. 
    
    \subsection{Advanced Packaging, 3D stacking, chiplets}  
    \label{subsection-chiplets}
     While Moore's Law has slowed down, we have found ways to continue the scaling towards lower process nodes (sub-10nm) using technologies like 3D stacking and Through-Silicon-via (TSV - a via being a vertical chip-to-chip connection) (Lim \cite{tsv-intro}), Near and sub Threshold Voltage (NTV) designs (Borkar et al. \cite{ntv-designs}), newer memory integration technologies, and more recently chiplets. Intel's Foveros (chiplets) \cite{intel-lakefield} is a new silicon stacking technique that allows different chips to be connected by TSVs so that the the cores, onboard caches/memory and peripherals can be manufactured as separate dies and can be connected together. By picking the best transistor for each function –  CPU, IO, FPGA, RF, GPU and accelerator – the system can be optimized for power, performance and thermals. Additionally, by stacking chiplets vertically Intel expects that it will be able to get around a major bottleneck in high-performance system-in-package design – memory proximity. While these technologies provide advanced packaging capabilities, cooling methods for such chips is currently a crucial area of development in the industry and will be an ongoing challenge.
	    


	
    
    
    \subsection{Thermodynamic computing, Landauer Limit and Quantum Computing}
	\label{subsection:thermodynamic-computing}
\begin{figure}[ht]
		\centering
		\includegraphics[width=\linewidth]{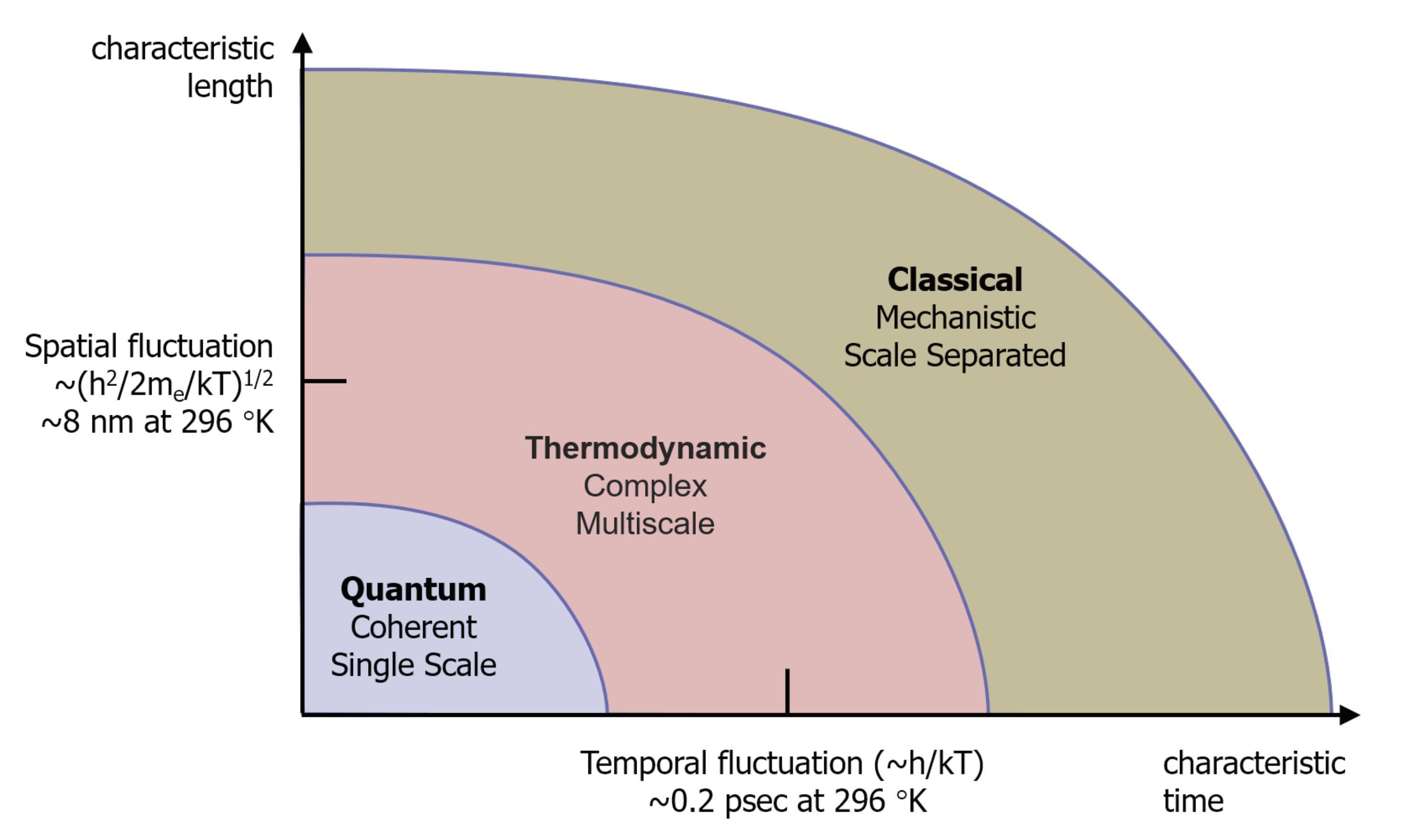}
		\caption{Comparing scales of classical, quantum and thermodynamic computing \cite{thermodynamic-computing}}
		\Description{Scales of computing}
		\label{fig:scales-of-computing}
	\end{figure}

	Richard Feynman, in his classic work \cite{feynman-computation} laid down the foundations of thermodynamic and quantum computing, which are now on the horizon. As detailed in the recent report on thermodynamic computing (Conte et al. \cite{thermodynamic-computing}), in today's "classical" computing systems that are based on transistors, quantum mechanical effects of sub-7/sub-5 nm are addressed by “averaging them” by appropriate tools and technologies. In such systems, components such as transistors are engineered such that their small-scale dynamics are isolated from one another. In the quantum computing domain, quantum effects are avoided by “freezing them” at very low temperatures. In the thermodynamic domain, fluctuations in space and time are comparable to the scale of the computing system and/or the devices that comprise the computing system. This is the domain of non-equilibrium physics and cellular operations, which is highly energy efficient. For example, proteins fold naturally into a low-energy state in response to their environment. The scale of these computing systems is shown in Figure \ref{fig:scales-of-computing}. In the figure, spatial and temporal fluctuation scales are estimated in terms of thermal energy (kT) and corresponding electronic quantum coherence times and lengths.
	
	Rolf Landauer, motivated by John von Neumann's considerations of entropy involved in computation, reasoned that when a bit of information is irreversibly transformed (erased, for example), or when two bits combine logically to yield a single bit (logic operations, for example), some information is lost, thereby resulting in a change in entropy of the system. \textit{Landauer's principle} \cite{landauer-limit} asserts that there is a minimum possible amount of energy required to erase one bit of information, known as the Landauer limit. Some recent work \cite{uc-berkely-nanomagnets} has demonstrated nanomagnetic logic structures that operate near the Landauer Limit, thereby raising the possibility of developing highly energy efficient computing systems in the future. 
	
	Quantum computing is another important architectural trend with different kinds of quantum hardware being built along with varying systems architectures, languages, runtime and workloads, as reported in Bertels et al. \cite{qc-delft} and Gyongyosi et al. \cite{quantum-computing-survey}. Getting such systems to work is the immediate focus across research and industry, and energy efficiency will be an important topic for the future. These topics are however, beyond the scope of this survey.

\section{Microarchitectural Techniques}
\label{section:microarchitectural-techniques}
	The fundamental techniques for energy efficiency involve fine-grained clock/power gating, dynamic frequency scaling (DFS) and dynamic voltage frequency scaling (DVFS). The basics of these techniques and thermal dissipation/management are described in exhaustive detail in Kaxiras and Martonosi \cite{martonosi-book-comp-arch-energy-eff}. Given the vast amount of work done in the area of energy efficiency techniques implemented in microarchitecture, we do not attempt to survey them all here. Instead, we focus only on those techniques that are visible and controllable by higher layers of the firmware/OS/software stack. 
	
	In this section, we focus on such microarchitectural techniques for energy efficiency across CPU, caches, memory and domain specific accelerators like GPUs and deep learning chips. 
	
	\subsection{Microarchitectural techniques for CPUs}
	\label{subsection:microarch-cpu}
	Power management for microprocessors can be done over the whole processor, or in specific areas. CPUs can have their execution suspended simply by stopping the issuance of instructions or by turning off their clock circuitry. Deeper power states successively remove power from the processor’s caches, translation lookaside buffers (TLBs), memory controllers, and so on. Deeper power states incure higher latency, and therefore extra energy is required to save and restore the hardware contents, or restart it. Modern processors support multiple low power states that can be exploited either by hardware (hardware idle detection) or through hints from the operating system scheduler based on heuristics such as next expected timer/interrupt, transition latency of different low power states, and current QoS setting dictated by other kernel components. As CPUs have evolved over the generations from single monolithic cores to multi-domain, multi-module and hybrid many core architectures, energy efficiency has been incorporated into different aspects. CPUs employ the following energy efficiency techniques:
    \begin{enumerate}
        \item \textit{Clock gating}: In this, the clock distribution to an entire functional unit in the processor is shutoff, thus saving dynamic (switching) power.
        \item \textit{Power gating}: Here, entire functional units of the processor are disconnected from the power supply, thus consuming effectively zero power.
        \item \textit{Multiple voltage domains}: Different portions of the chip are powered by different voltage regulators, so that each can be individually controlled for DVFS scaling power gating. Recent designs use on-die and on-chip voltage regulators that can do fine-grained power management through CPU microcode or low level firmware (Haj-Yahya et al. \cite{isqed-2019-powerdelivery}).
        \item \textit{Multi-threshold voltage designs}: Different transistors in the design use different threshold voltages to optimize delay and/or power (Hemantha et al. \cite{multithreshold-cmos}).
        \item \textit{Dynamic frequency scaling (DFS)}: The clock frequency is adjusted statically or dynamically to achieve different power/performance trade-offs.
        \item \textit{Dynamic voltage scaling (DVS)}: The supply voltage of the processor is adjusted statically or dynamically to achieve different power/performance and reliability trade-offs.
        \item \textit{Dynamic voltage and frequency scaling (DVFS)}: Both voltage and frequency are varied dynamically to achieve better power/performance trade-offs than either DFS or DVS alone can provide.
    \end{enumerate}

    Beyond the CPU cores, uncore components like caches, translation lookaside buffer and others, also implement energy efficiency techniques as embedded microprocessors devote nearly 40\% of their power budget to uncore/caches. Current cache implementations use several techniques. Smart sizing caches is done by the micro code in the processor core. In Varadarajan et al. \cite{cache-molecules}, the authors define application specific cache partitions, called \textit{cache molecules}, that are resized to address performance targets for applications. Some other examples include drowsy caches, dynamic clock gating based on operand width and instruction compression, among others; these are detailed in the book Kaxiras and Martonosi \cite{martonosi-book-comp-arch-energy-eff}.


	\subsection{Microarchitectural techniques for Memory}
	\label{subsection:microarch-mem}
	Memory technology has evolved across DDR3/4/5, LPDDR, and more recently non-volatile memory (NVM) and these have enabled different levels of performance and power management with features such as clock frequency control and varying degrees of shallow/deep self-refresh. Newer memories like non volatile memory (NVM) exhibit different power and energy efficiency characteristics across reads, writes and self-refresh states. Recent system design, application, and technology trends that require more capacity, bandwidth, efficiency, and predictability out of the memory system make the memory system an important system bottleneck. At the same time, DRAM and flash technologies are experiencing technology scaling challenges that make the maintenance and enhancement of their capacity, energy efficiency, performance, and reliability significantly expensive with conventional techniques. 
	
	
	Energy efficiency in memory is important in the context of workloads like deep neural networks (DNNs). System designs that enable accelerated processing of DNNs with improved energy efficiency but without trading off accuracy or increasing hardware costs have become indispensable. Computing of such applications is governed by data movements rather than the execution of algorithmic or logical functions. Hence, dependence of system performance on the efficiency of processor-memory interaction is seeing an all-time high as we have striven to push beyond the memory wall (Radulovic et al. \cite{intro1}, McKee \cite{intro2}). With memory technologies like 3D-stacked memories (Liu et al. \cite{intro3}) and non-volatile memories (Zhang et al. \cite{intro4}), the \textit{memory wall} issue is being addressed to some degree. However, the high bandwidth and greater storage capacity of such alternatives to conventional DDR systems for main memory can be helpful only if they are intelligently utilized by the system. This requires a synergy of the resource requirement of the workload with the available bandwidth, parallelism and data access hierarchy of the underlying memory system via hardware-software techniques. Micron's Hybrid Memory Cube (HMC) has made a compelling case for realization of a high throughput and low energy solution for massively parallel computations with their extensive bandwidths (Pawlowski \cite{hmc2}) facilitated by through silicon via (TSV) technology (Lim \cite{tsv-intro}) and near-data processing (NDP) (Balasubramonian et al. \cite{intro14}) in the logic layer. An apt architectural design of memory layers as well as the logic layer of HMC can enable the effective bandwidth to be as close as possible to the maximum available bandwidth (Hadidi et al. \cite{hmc1}), Radulovic et al. \cite{intro12}).
	
	Some systems use partial array self-refresh (PASR), where memory is divided into banks, each of which can be powered up/down independently. If any of those banks of memory are not needed, that memory (and its self- refresh mechanism) can be turned off. The result is a reduction in power use, but data stored in the affected banks is also lost. Correspondingly, this requires operating system support for intelligent memory allocation. 
	
	\subsection{Microarchitectural techniques for GPUs}
	\label{subsection:microarch-gpu}
	Modern GPUs consume a significant amount of power - anywhere from \~50-300W (or even more). However, GPUs provide better performance-per-watt than CPUs for specific workloads. The techniques for improving energy efficiency of GPUs largely overlap with those used for CPUs, with some variations and additions. A detailed survey is presented in Mittal et al. \cite{mittal-vetter-ee-gpu} and some of the key techniques are highlighted here:
	    \begin{enumerate}
	        \item \textbf{GPU DVFS}: Many current GPUs have separate clocks and voltage domains, thereby making them ideal candidates for clock/frequency scaling, voltage scaling, or both through hardware/software orchestration. Typically, in low power GPUs (in handhelds, for example), the chip is divided into three power domains - vertex shader, rendering engine, and RISC processor, and DVFS is individually applied to each of the three domains, thereby allowing for finer orchestration of the power domains.
	        \item \textbf{CPU-GPU orchestration}: Instead of using a single GPU with each CPU, using multiple GPUs with each CPU enables achieving speedup in execution time and improving the usage of the CPU, thereby improving the energy efficiency of the system. Further, since during the execution of the CUDA kernel the host CPU remains in the polling loop without doing useful work, the frequency of the CPU can be reduced for saving energy while ensuring that CPU frequency is optimal for the bus between the CPU and GPU. Since the range of CPU frequencies is generally larger than that of the bus, CPU frequency can be scaled without affecting GPU performance. Also, for specific workloads, using CPU DVFS can be employed while it stays in busy-waiting for the GPU to complete computations, thereby achieving energy savings with little performance loss. Most of these can be orchestrated through hardware and software components.
	        \item \textbf{Energy efficiency in GPU components}: GPU components such as caches, global memory, pixel and vertex shader can all be managed through dynamic clock and power gating. Since GPUs employ a large number of threads, storing the register context of these threads requires a large amount of on-chip storage. Also, the thread scheduler in the GPU needs to select a thread to execute from a large number of threads, access large register files, etc. which consumes substantial energy. Similarly, instruction pipeline, shared registers, last-level caches can also be made more energy efficient through hardware and microarchitectural techniques.
	    \end{enumerate}
	 
	\subsection{Microarchitectural techniques for AI accelerators}
	\label{subsection:microarch-accel}
	An AI accelerator chip has three main elements — a large amount of data, algorithms to process the data (configurable by software), and the physical architecture where data processing/calculation is carried out. Such accelerators tend to have regular architectures - large arrays with hundreds or thousands of processors, arranged in clusters repeated across the chip and consuming power in the order of tens or even hundreds of watts. The key energy efficiency techniques for such chips comprise of hardware/software partitioning of the workload, mapping of data structures into on-chip and off-chip memory, grouping of components into power domains, power management policy (race-to-halt typically), and enter idle states when parts of the chip are idle. Designs typically also include many temperature sensors across the die — for example, one per processing cluster, to aid in aggressive thermal management.

	Given the data-intensive nature of CNN algorithms (ML performance and power is dominated by data movement, not compute), several implementations have looked at accelerating the memory subsystem. Recent works like Tetris (Gao et al. \cite{bg3}), Neurostream (Azarkhish et al. \cite{bg4}), Neurocube (Kim et al. \cite{bg5}) have proposed CNN accelerator implementation in the logic layer of Hybrid Memory Cube (HMC). Here, in order to alleviate the bandwidth pressure on the data-path between the processor chip and the main memory chip, and to get rid of the large on-chip local memory that occupy more than 50\% of the chip (Eyeriss - Chen et al. \cite{eyeriss-dataflow}), an array of processing elements and register files (as and where needed) are incorporated in the logic layer of the 3D-stacked DRAM module. Azarkhish et al. \cite{bg4} use HMC as a co-processor for CNN acceleration through synchronization free parallelism while Kim et al. \cite{bg5} embed Neurocube, which are specialized state-machines within the vault controllers of HMC to drive data into the processing elements in the logic layer. Some accelerators use strategies such as optimized memory use and the use of lower precision arithmetic to accelerate calculations and increase throughput of computation, however, they tend to be designed for specific use cases and markets. Most of the accelerators support traditional clock and power gating; some of them support DFS / DVS / DVFS, making them amenable to standard energy efficiency algorithms through hardware software orchestration. 

    The data-intensive nature of CNN algorithms is in contrast with von Neumann execution models and this has motivated non-von Neumann models of computation like dataflow, spiking neural networks, and other forms of brain-inspired computing. Chen et al. \cite{eyeriss-dataflow} propose Eyeriss, an optimized algorithmic dataflow for CNNs by exploiting local data reuse and optimization of intermediate data movement. Tetris (Gao et al. \cite{bg3}) uses the dataflow model of Eyeriss along with scheduling and partitioning in software to implement CNN acceleration in HMC. In Farabet et al. \cite{farabet2011neuflow}, the authors present a compiler that transforms high level dataflow graphs into machine code representations. Another work SmartShuttle (Li et al. \cite{li2018smartshuttle}) adaptively switches among different data reuse schemes and the corresponding tiling factor settings to dynamically match different convolutional layers. Its adaptive layer partitioning and scheduling scheme can be added on existing state-of-the-art accelerators to enhance performance of each layer in the network. The industry has also seen some innovative products in this space. Wave Computing \cite{nicol2017coarse} presents an implementation of a dataflow architecture as an alternative to train and process DNNs for AI especially when models require a high degree of scaling across multiple processing nodes. Instead of building fast parallel processors to act as an offload math acceleration engine for CPUs, Wave Computing's dataflow machine directly processes the flow of data of the DNN itself. Energy efficiency of deep learning accelerators is covered in more detail in Ganguly et al. \cite{Ganguly2019TowardsEE}. 
    

	\section{Specification}
	\label{section:specification}
\begin{figure}[ht]
		\centering
		\includegraphics[width=\linewidth]{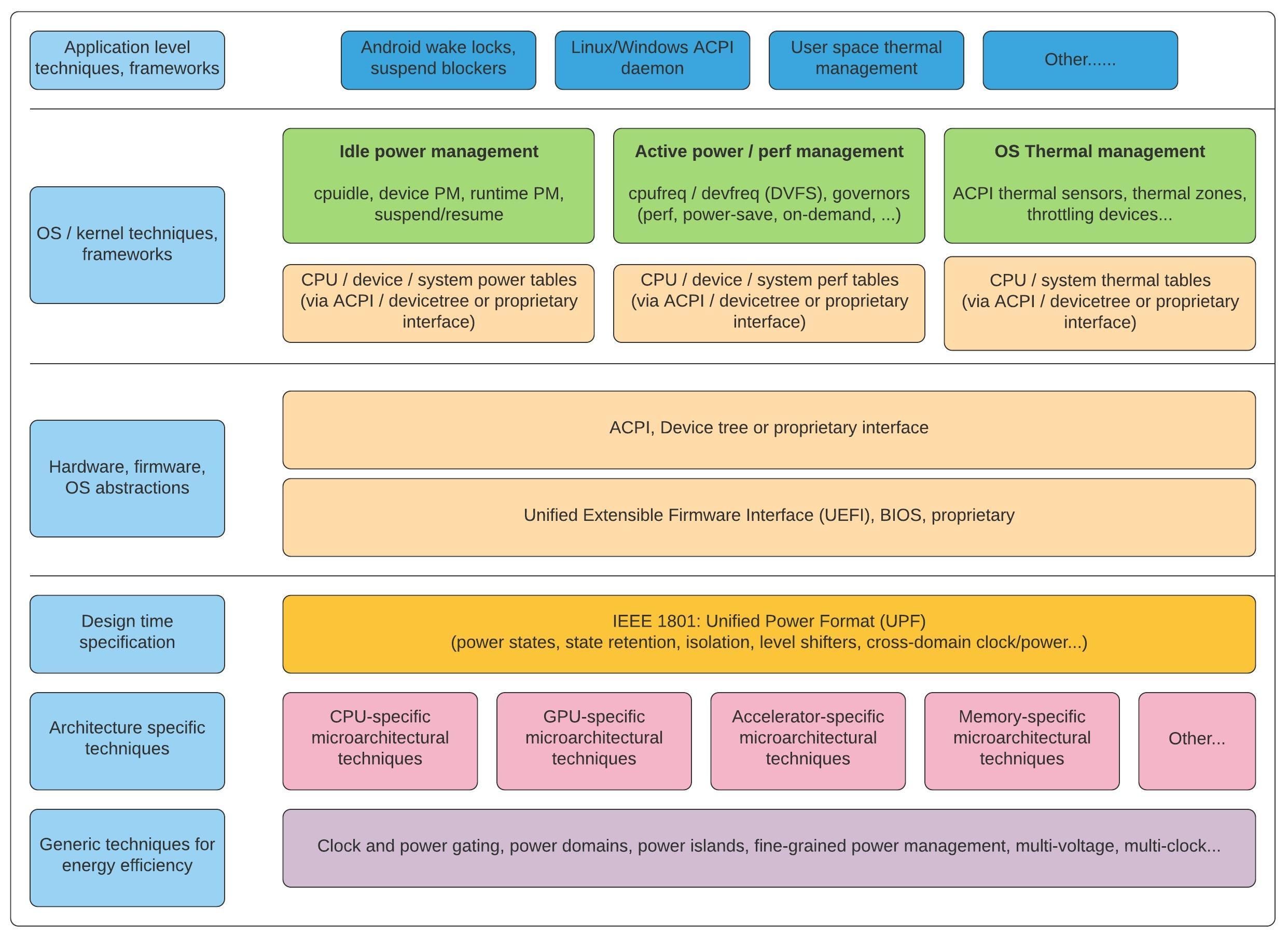}
		\caption{Specifications and abstractions at different levels} 
		\Description{Low Power specifications and abstractions}
		\label{fig:low-power-specs-abstractions}
	\end{figure}

	Energy efficiency techniques at hardware / RTL level (clock gating, multi-voltage design, power gating and DVFS) are specified using industry standards like IEEE 1801 Unified Power Format (UPF). At the microarchitectural level, techniques described in Section \ref{section:microarchitectural-techniques} are used and are specified using proprietary methods. At the hardware-firmware-OS level, a different set of specifications are used to describe underlying hardware, power, performance and thermals. Further up the stack, the OS and applications use these abstractions to implement various energy efficiency techniques, such as the Linux Idle and Runtime PM framework, DVFS governors, thermal management algorithms and Windows Connected Standby. The specifications and abstractions used at, and across, each levels are now described and are illustrated in Figure \ref{fig:low-power-specs-abstractions} (the different colours are to delineate different layers and components).
	
	\subsection{Accellera Unified Power Format}
	\label{subsection:accellera-upf}
	In 1991, Open Verilog International and VHDL International were formed to encourage open collaboration, portability and interoperability in electronic design automation (EDA). Recognizing the need for a wider impact across the whole industry of IP designs, vendors, integrators, ODMs/OEMs, the Accellera Systems Initiative was formed in 2000 as a merger of Open Verilog International and VHDL International. The goal was to build a standards organization to support interoperability and open interfaces for EDA and IC design/manufacturing and testing \cite{accellera/upf}. Accellera then merged with the SPIRIT Consortium, which was a standards group composed of vendors and users of EDA tools focusing on SOC level information. SPIRIT stood for "Structure for Packaging, Integrating and Re-using IP within Tool-flows". The SPIRIT Consortium had defined IP-XACT, an XML schema for vendor-neutral description of design components, and SystemRDL, a language for describing registers in components. Accellera formalize several other EDA standards such as:
	\begin{itemize}
	    \item Universal Verification Methodology (UVM)
        \item Open Verification Library (OVL)
        \item Standard Co-Emulation Modeling Interface (SCE-MI)
        \item Unified Coverage Interoperability Standard (UCIS)
        \item IP-XACT - Update of IEEE 1685 and Recommended Vendor Extensions
        \item Intellectual Property (IP) Tagging
        \item SystemC
        \item SystemRDL
        \item Open Core Protocol (OCP)
	\end{itemize}
	Accellera is less constrained than the (IEEE) and is therefore the starting place for many standards, interest groups, study groups to evaluate different ideas in the areas of EDA. Once mature and adopted by the broader community, the standards are usually transferred to the IEEE. The Unified Power Format (UPF) or the IEEE 1801 standard was born in this manner. A Unified Power Format technical committee was formed by the Accellera organization, and UPF 1.0 was approved and published and this was donated to the IEEE as a basis of this standard in 2006.

	\subsection{IEEE 1801: Unified Power Format}
	\label{subsection:upf}
	The microarchitectural techniques for energy efficiency translate to hardware through a some important concepts at the RTL or lower levels:
	\begin{enumerate}
	    \item \textbf{Power domains}: These are independently powered domains, enabling the application of different power reduction techniques in each domain.
	    \item \textbf{State retention}: It is important to save essential state when power is off, and to restore it when the power is turned back on. For this, special state-retention elements can be added to keep a minimal amount of power available to registers whose contents must be preserved during power shutdowns.
	    \item \textbf{Isolation}: This is to ensure correct logical and electrical interactions between domains belonging to different power states. To do this, a tool can insert isolation cells on signals coming from regions that are turned off.
	    \item \textbf{Legal power states}: Only legal power state transitions must be allowed across components.
	    \item \textbf{Level shifters}: To ensure communication between domains powered by different voltage levels, level shifters are added to signals crossing between regions with different voltages and different switching thresholds.
	\end{enumerate}
	Across all these techniques, it is crucial to have a common, unambiguous representation of low power design intent across designers, verification engineers, design and verification tools.
	
	IEEE 1801 Standard for the Design and Verification of Low Power Integrated Circuits, also called the Unified Power Format (UPF), is a standard for specifying the power intent and low power methods in early phases of design. UPF allows for specifying hardware systems with power as a key consideration and UPF scripts help describe power intent, or power management constructs / features. For example - which power rails are to be routed to individual blocks, when are blocks expected to be powered up or shut down, how voltage levels should be shifted between two different power domains and the type of measures taken for retention registers if the primary power supply to a domain is removed. Additionally, specifying power features in a standard format allows for several design and verification tools to validate the complex design. Beyond the obvious importance of using standardized formats across all phases of design, the other importance of using UPF arises from the fact that often large blocks of hardware IP are re-used either in different systems-on-chip designs or several different generations of a particular system or even for porting a proven system to a different target technology. This is, therefore, a particularly important problem for hardware IP suppliers who need to be able to supply descriptions of power intent for products to their customers without having any information about what implementation-specific decisions might be taken by the customer, or how their IP is integrated into a different hardware / SOC design.

	The latest standard, UPF 3.0, released in 2016, has improved capabilities for adding bottom-up implementation flow, power models, and high-level power analysis. The ability to develop energy-efficient platforms, including the hardware, software and system power management components of the platform, requires the ability to use appropriate levels of design abstraction for the task at hand. With UPF 3.0, architects can now model the salient power related characteristics of a piece of IP for use at the system level, thereby providing a foundation for building complex system level power models in a standardized manner. Using UPF-based hardware designs as reusable components in other SOCs is an important area for power/performance projections using power models of individual hardware comopnents leading up to system level power models. This is an important area of cross industry collaboration and standardization in the IEEE P2416 \cite{ieeep2416} working group. 
	
	

	\subsection{ACPI and DeviceTree}
	\label{subsection:uefi-acpi-devtree}
	UPF is a design time specification for low power and it is disconnected from runtime management by system software. Over the years, several proprietary and industry consortiums have attempted to define abstractions for runtime management, which we will now describe.

	\subsubsection{Advanced Configuration and Power Interface (ACPI)}
	\label{subsubsec:acpi}
	ACPI \cite{uefi-acpi} is a standard for runtime management of hardware. The scope of ACPI comprises system run-time configuration, power and thermal management as well as hardware error handling support. ACPI is, essentially, a standardized way to enable the operating system to discover, configure and initialize the system's hardware. It provides runtime tables for power management (among other things) - power states supported by the CPU(s), CPU hierarchy, DVFS states supported and associated transition latencies, thermal sensors supported on the platform, thermal states supported and thermal throttling order. The important thing to note is that UEFI is not tied to ACPI and will work with any firmware description. Similarly, ACPI does not depend on UEFI, and can work with any other low level device initialization framework as well such as U-Boot or BIOS. ACPI is a very active industry working group and is constantly being updated and this is the primary OS power management technique used across different segments of computing - laptops, desktops, HPC, data center systems and supercomputers.
	
	\subsubsection{Device Tree (DT)}
	\label{subsubsec:devicetree}
	While ACPI was historically created for x86 platforms, the ARM ecosystem developed Device Tree to describe the same information for ARM-based devices. Thus, ACPI and DT overlap in that they both provide mechanisms for enumerating devices, attaching additional configuration data to devices (which can be used by higher layers of software). Rafael Wysocki \cite{acpi-devtree-wysocki} goes into details of the commonalities between ACPI and Device Tree and the convergence between the two standards.
	
	The biggest difference between DT and ACPI is that DT is effectively a somewhat structured mechanism for passing arbitrary data, while ACPI tends to provide standardised data. 
	
	

\section{Modeling and Simulation}
\label{section:modeling-simulation}
The main goal of simulation is to model new research ideas for parts of a system (processor, memory, accelerator and others) or a complete system (SOC or server) and estimate metrics such as performance and energy. While initial generation of tools catered to building functional, timing/cycle-accurate models for performance estimation, subsequent tools incorporated power, energy and thermal modeling, simulation and estimation/projections and also the ability to run real, or close-to real workloads as well as full operating systems. Some key modeling/simulation tools across different kinds of hardware are illustrated in Table \ref{tab:modeling-simulation}. In this section, we focus primarily on power, energy and thermal modeling/estimation tools for multicore processors, domain-specific accelerators, and SOC/full chip systems.
	
	
   \begin{table}
    	\caption{Summary of Modeling and Simulation tools}
		\label{tab:modeling-simulation}
    \begin{tabular}{|c|l|}
        \toprule
        \textbf{Domain} & \textbf{Key work, surveys or books} \\
        \midrule
        \multicolumn{1}{|m{6cm}}{Processor and multiprocessor simulators} & 
		\multicolumn{1}{|m{6cm}|}{gem5 \cite{gem5}, Multi2sim \cite{multi2sim}, MARSSx86 \cite{marss-x86-simulator}, PTLsim \cite{PTLsim} and ZSim \cite{zsim}, Akram et al. (\cite{ayaz-lina-2016}, \cite{lina-power-perf-ISAs}), Eeckhout \cite{eeckhout-processor-perf-methods}} \\ \hline
		
		\multicolumn{1}{|m{6cm}}{Cache Simulators} & 
		\multicolumn{1}{|m{6cm}|}{gem5 \cite{gem5}, CACTI \cite{cacti-3.0-cache-power}, Brais et al. \cite{ppp-cache-simulators}}\\ \hline
		\multicolumn{1}{|m{6cm}}{Memory Simulators} & 
		\multicolumn{1}{|m{6cm}|}{DRAMPower \cite{dram-power-tool}, DRAMSim2 \cite{dramsim2}, VAMPIRE \cite{memory-sim/vampire}, Ramulator \cite{ramulator},  NVMain \cite{memory-sim/nvmain}, NVM Streaker {memory-sim/nvmstreaker}, DRAMSim3 \cite{memory-sim/dramsim3}} \\ \hline
		
		\multicolumn{1}{|m{6cm}}{GPU Simulators} & 
		\multicolumn{1}{|m{6cm}|}{GPUWattch \cite{gpuwattch}, GPGPU-Sim \cite{gpgpu-sim}, MGPUSim \cite{MGPUSim}, AccelWattch \cite{accelwattch-gpu-simulator}, Bridges et al. \cite{gpu-power-simulation-modeling}}\\ \hline
		
		\multicolumn{1}{|m{6cm}}{Accelerator Simulators} & 
		\multicolumn{1}{|m{6cm}|}{Alladin \cite{aladdin-accel-sim}, Minerva \cite{minerva-dnn-sim}, FireSim \cite{firesim}, Akram et al. \cite{ayaz-lina-2016}}\\ \hline
			
		\multicolumn{1}{|m{6cm}}{SOC and full system simulators} & 
		\multicolumn{1}{|m{6cm}|}{PARADE \cite{parade-full-sys-sim}, gem5 \cite{gem5}, McPAT \cite{mcpat}, SoftSDV \cite{intel-softsdv}}\\ \hline
			
		\multicolumn{1}{|m{6cm}}{Power and Energy Simulators} & 
		\multicolumn{1}{|m{6cm}|}{Wattch \cite{wattch-power}, SimplePower \cite{simple-power}, IBM PowerTimer tool \cite{ibm-powertimer}, McPAT \cite{mcpat}, PowerAnalyzer \cite{power-analyzer}, FPGA Simulators (Anderson et al. \cite{fpga-power})} \\ \hline
		
		\multicolumn{1}{|m{6cm}}{Power Delivery Simulators} & 
		\multicolumn{1}{|m{6cm}|}{VoltSpot \cite{Voltspot}} \\ \hline
	
		\multicolumn{1}{|m{6cm}}{Thermal Simulators} & \multicolumn{1}{|m{6cm}|}{Kaxiras and Martonosi \cite{martonosi-book-comp-arch-energy-eff}, TEMPEST \cite{tem2p2est}, Hotspot \cite{hotspot}, SESCTherm \cite{SESCTherm}, Power Blurring \cite{power-blurring}, Intel Docea \cite{intel-docea}, Sultan et al. \cite{sarangi-thermal-survey}} \\ \hline
        \bottomrule
    \end{tabular}
    \end{table}
    
	

\subsection{Processor and full system Simulators}
\label{subsection:processor-sim}

The first processor power simulators, such as Wattch \cite{wattch-power}, were introduced around 2000, with the last decade seeing more multiprocessor / hetero-core simulators, complete with OS and runtime system so that entire workloads can be simulated.
 
The book by Eeckhout \cite{eeckhout-processor-perf-methods} details the state-of-the-art in computer architecture performance evaluation. The book focuses on fundamental concepts and ideas for obtaining accurate performance data and covers various topics in performance evaluation. Some of the most popular x86 processor simulators used currently are: gem5 \cite{gem5}, Multi2sim \cite{multi2sim}, MARSSx86\cite{marss-x86-simulator}, PTLsim \cite{PTLsim} and ZSim \cite{zsim}. gem5 \cite{gem5} is an event-driven full-system simulation tool, which is extensively used in both academia and industry. It is an event driven simulator, and can also keep track of events on a cycle-by-cycle basis, which makes its accuracy comparable to a cycle-level simulator. It supports many instruction set architectures (ISA)s: ARM, x86, MIPS, SPARC, ALPHA, Power and RISC-V. Multi2Sim \cite{multi2sim} is a simulator that mainly targets GPUs and simulates CPU-GPU architectures. It supports many ISAs - x86, MIPS, ARM and AMD Evergreen ISA. MARSS \cite{marss-x86-simulator} is an open source, full system simulation tool built on QEMU \cite{qemu}, to support cycle-accurate simulation of homogeneous and heterogeneous multicore x86 processors. It includes detailed models of coherent caches, interconnections, chipsets, memory and IO devices. It also simulates the execution of all software components in the system, including unmodified binaries of applications, OS and libraries. PTLsim \cite{PTLsim} is a cycle-level simulator that has the ability to simulate complete OS using Xen hypervisor. It makes use of co-simulation and is capable of modeling a superscalar out-of-order core. ZSim \cite{zsim} is a parallel application-level timing simulator for x86-64 architectures. It focuses more on simulating memory hierarchies and many core heterogeneous (single-ISA) systems. It supports modeling both out-of-order (OOO) and in-order (IO) pipelines. Akram and Sawalha \cite{ayaz-lina-2016} discuss these in detail and Akram et al. \cite{lina-power-perf-ISAs} discuses power and performance comparisons of different processor architectures and ISAs.

Wattch \cite{wattch-power} was one of the first tools to provide accurate power estimation of processors. It developed a framework for analyzing and optimizing microprocessor power dissipation at the architecture-level thereby allowing architects to make high-level analysis of power tradeoffs.  SimplePower \cite{simple-power} was introduced as a means of doing detailed whole processor analysis of dynamic power. It focused on in-order five-stage pipelines, with detailed models of integer ALU power as well as other regions of the chip. The Wattch tool built on cache modeling from Cacti \cite{cacti-3.0-cache-power}, and provided parameterized activity factor-based estimates as well. Both SimplePower \cite{simple-power} and Wattch \cite{wattch-power} were both based on analytic power modeling techniques. The IBM PowerTimer tool \cite{ibm-powertimer} provides a processor simulator based on empirical techniques — one can estimate the power consumption of a particular architectural module by using the measured power consumption in an existing reference processor, and applying appropriate scaling techniques for design and process technology. This tool thereby allows architects to estimate power of future generation designs early in the design phase. McPAT \cite{mcpat} can simulate timing, area and power of multicore processors. PowerAnalyzer \cite{power-analyzer} is a power evaluation tool suitable for calculating power consumption for complete computer systems. Power consumption of FPGAs is also an important area, hence modeling the power consumption of FPGA-based systems has also gained importance in recent years. Anderson et al \cite{fpga-power} provide a survey of power estimation techniques for FPGAs. The authors formulate empirical prediction models for net activity for FPGAs.

In addition to such architectural power simulators, the other important area is the simulation of the on-chip power delivery system itself. With the end of Dennard's scaling, as transistor densities increased, threshold and supply voltages could no longer decrease fast enough to prevent an exponential growth in on-chip power densities (Mack \cite{50-years-moores-law}). With the continued growth of tighter device integration, sophisticated power delivery networks (PDNs) are required to not only deliver sufficient current to switching transistors, but also to remove the heat generated by silicon chips. A modern PDN usually consists of several voltage regulator modules (VRM) and decoupling capacitors. VoltSpot \cite{Voltspot} is an architecture-level model of the on-chip power delivery network that can be integrated with a performance simulator (such as Gem5) and power estimation tool (such as McPAT), thus providing architects with the tools necessary to explore the effect of PDN design, exploration of run-time IR drop and noise prediction, avoidance, and mitigation. VoltSpot Version 2.0 extends VoltSpot's modeling capability to cover 3D-ICs and Through-Silicon-Vias (TSV) as well. Vaisband and Friedman \cite{vaisband-pdn}, the authors introduce the concept of power network-on-chip (PNoC) as a general scalable platform for modern power delivery networks. PNoCs form the foundation of modern on-chip power delivery for SOCs to enable enhanced power control and real-time management of resource sharing for scalable management of heterogeneous integrated circuits.

\subsection{Cache Simulators}
\label{subsection:cache-sim}

The memory technology used in a cache determines the power, performance and reliability of the cache. SRAM is typically used to build caches, but DRAM has also been used for last-level caches in processors. Most cache simulators have no inherent notion of memory technology; rather they take the characteristics of the desired technology as inputs. Newer memory technologies such as non-volatile caches and 3D caches can be abstracted similarly. CACTI 3.0 (Shivakumar and Jouppi \cite{cacti-3.0-cache-power}) is one of the most popular cache simulators - it integrates cache access time, cycle time, area, aspect ratio, and power model. By integrating all these models together users can have confidence that tradeoffs between time, power, and area are all based on the same assumptions and hence are mutually consistent. Area, Power and Timing (APT) models are usually integrated into the simulator. McPAT \cite{mcpat} includes the Cacti-P modeling tool for SRAM, DRAM, and 3D stacked DRAM caches. gem5 \cite{gem5}, Multi2Sim \cite{multi2sim} are available with integrated McPAT models. From a power/energy perspective, simulators should also provide support for dynamic voltage and frequency scaling (DVFS). This involves maintaining separate clock domains and capturing the interactions between them. Many simulators like gem5 \cite{gem5} provide support for DVFS. Brais et al. \cite{ppp-cache-simulators} provide a detailed discussion on 28 CPU cache simulators, including recent simulators.
	
\subsection{Memory Simulators}
\label{subsection:memory-sim}
Due to the constant gap between processor and memory speeds, evaluating memory-system designs before they are implemented in hardware is extremely important. Older techniques using analytical methods \cite{agarwal-cache-simulator} and trace-driven memory simulation \cite{Uhlig2000} to predict memory system performance have now been replaced with accurate simulators of real memory systems/controllers.


Newer memory simulation techniques have largely replaced trace driven techniques. DRAMPower \cite{dram-power-tool} is an open source tool for fast and accurate DRAM power and energy estimation for DDR2/DDR3/DDR4, LPDDR/LPDDR2/LPDDR3 and Wide IO DRAM memories. The tool can be employed at both command level and transaction level. Users employing DRAM memory controllers in their existing system setup can log the DRAM command traces and employ DRAMPower at the command-level. Users without access to DRAM memory controllers can make use of the optional DRAM command scheduler, which dynamically schedules and logs DRAM commands, corresponding to the incoming memory transactions, as if it were a regular memory controller. The generated DRAM command schedule is analyzable for real-time applications. DRAMSim2 \cite{dramsim2} is a cycle-accurate DDR2/DDR3 simulator. In order to effectively model the dynamics of CPU and memory subsystems, DRAMSim2 is integrated with MARSSx86 \cite{marss-x86-simulator}, a full x86 system simulation environment. The default fixed memory latency in MARSSx86 is replaced with calls to add requests to DRAMSim2; a callback function within MARSSx86 sends these requests back through the cache hierarchy to the CPU when they are complete. Ghose et al. \cite{memory-sim/vampire} observed that state-of-the-art DRAM power models are often highly inaccurate, as these models do not reflect the actual power consumed by real DRAM devices. The authors conducted a comprehensive experimental characterization of the power consumed by 50 modern real-world DRAM (DDR3L) modules and based on the observations, developed VAMPIRE - Variation-Aware model of Memory Power Informed by Real Experiments \cite{memory-sim/vampire}. VAMPIRE is a new, accurate power consumption model for DRAM that takes into account module-to-module and intra-module variations, and power consumption variation due to data value dependency. 


Given the rapid adoption of new memory technologies such as GDDR5, High Bandwidth Memory (HBM), Wide IO 1/2 as well as others that are in research phase, there is a growing need for an extensible DRAM simulator that can be used to model many different memory systems. Ramulator \cite{ramulator} is a fast and cycle-accurate DRAM simulator that is built from the ground up for extensibility. Unlike existing simulators, Ramulator is based on a generalized template for modeling a DRAM system, which is only later infused with the specific details of a DRAM standard. Thanks to such a decoupled and modular design, Ramulator is able to provide out-of-the-box support for a wide array of DRAM standards: DDR3/4, LPDDR3/4, GDDR5, WIO1/2, and HBM. It is also released as an open source tool under BSD license. 3D packaging of DRAM and the integration of CPU and DRAM on the same die allows for higher density,  better performance and also lower power consumption. However, accurate simulation tools have not kept up with DRAM technology, especially for the modeling of 3D DRAMs. DRAMSim3 \cite{memory-sim/dramsim3} is a cycle accurate DRAM simulator that offers thermal modeling along with performance modeling.  

In order to simulate emerging non-volatile memory (NVM) technologies such as PCRAM and STT-RAM, NVMain was developed. NVMain \cite{memory-sim/nvmain} is an architectural level simulator that can model memory design with both DRAM and emerging non-volatile memory technologies. Similarly, NVM Streaker \cite{memory-sim/nvmstreaker} is a fast and reconfigurable simulator and it simulates NVM access costs using disturbed DRAM accesses and commonly configurable hardware parameters.

\subsection{GPU Simulation}
\label{subsection:GPU-sim}
	
Architecturally, modern GPUs contain anywhere from a few dozens to several thousands of small processors called streaming processors (SPs). Depending on the GPU specific architecture, \~8 to 64 SPs are organized into a streaming multiprocessor (SM) along with a few special function units (SFUs), which handle the more complex math operations. It is important to note that SMs do not have a branch unit, unlike CPUs. Each SM includes a multi-thread instruction fetch and issue unit, a L1 cache as well as a shared L2 cache shared by all SMs. GPU’s complex internal memory hierarchy and thousands of processors make them challenging candidates for power modeling. Internal hardware events were not observable in earlier generations, this has changed in recent architectures. 

GPUWattch \cite{gpuwattch} is a tool for modeling the power consumption of GPU architectures, but it was designed to model (and was validated against) older architectures with fewer energy efficiency optimizations. As reported in \cite{accelwattch-gpu-simulator}, attempts to model recent GPUs such as NVIDIA's Pascal, Volta and Turing using the methodology employed by GPUWattch produces significant inaccuracies, both in terms of absolute numbers and in terms of the relative power consumption of individual hardware components. Event-driven cycle-accurate simulators such as GPGPU-Sim \cite{gpgpu-sim}, Multi2Sim \cite{multi2sim} and MGPUSim \cite{MGPUSim} have been used for GPU architecture research. All of these also lacked the ability to power model native machine ISA. AccelWattch \cite{accelwattch-gpu-simulator} is a recent GPU power modeling tool that is configurable, capable of cycle-level calculations in emulation and trace-driven environments, and supports DVFS. AccelWattch is the only power model capable of modeling both virtual ISA and native machine ISA instructions, and is the only open-source tool capable of modeling closed-source workloads — it only needs a binary. In addition,AccelWattch is perhaps the only GPU power model that can be driven by either pure software performance models (e.g., Accel-Sim [20]), or hardware performance counters commonly found in modern GPUs (thereby capturing execution on real silicon), or a combination of the two. These AccelWattch variants allow researchers to balance the trade-off between power model accuracy and performance modeling effort. Bridges et al. \cite{gpu-power-simulation-modeling} present a detailed survey of GPU power and performance estimation and modeling across different GPU architectures, estimation and projection methodologies.

\subsection{Thermal Modeling}
\label{subsection:thermal-sim}
	
The ability to model thermal behavior is important especially for small form factor devices like smartphones and handhelds where the heat flows are critical in determining the usage of the device (and restrictions therein). Thermal modeling is also heavily used in large server farms and data centers to be able to administratively monitor and manage load across servers. Thermal modeling has several aspects ranging from designing thermals for a microprocessor alone to provisioning thermal sensors, and cooling of larger systems or data centers. In the past, the focus was on CPU thermal modeling, estimation and analysis; the focus has now moved to platform level thermal modeling, estimation and control mechanisms. Kaxiras and Martonosi \cite{martonosi-book-comp-arch-energy-eff} describe in detail the relationship between power and temperature and show the exponential dependence of power on temperature and the cyclic relationship — thermals depend on power dissipation and density; on the other hand, power also depends on temperature. 

$TEM^2P^2EST$ \cite{tem2p2est} was one of the first thermal models, where temperature was modeled based on power dissipation and density values. It is a flexible, cycle-accurate microarchitectural power and performance analysis tool based on SimpleScalar \cite{simplescalar}. The simulator generates power estimates based on either empirical data or analytical models and supports dynamic and leakage power and process technology scaling options as well as effects of clock throttling. The main drawback was that it modeled only the CPU, but not other regions or other architectural units. Skadron et al. \cite{hotspot} proposed and validated the HotSpot approach, a compact RC model for localized heating in high-end microprocessors. This was a complex model that considered both the lateral relationships between units on chip, as well as the vertical heating/cooling relationships between the active portion of the silicon die and the attached heat spreader and heat sink layers that seek to even out temperature and draw heat away from the active silicon. In recent SoCs, thermal modeling has taken up even higher prominence given that some of these smaller devices have no active cooling mechanisms like fans. Platform architects build hardware prototypes with heat generators that are modeled on actual physical components, and then test the prototypes in thermal chambers to analyze heat flow. SESCTherm \cite{SESCTherm} is a novel temperature modeling infrastructure that offers accurate thermal characterization. This framework is based on finite difference methods and equations. Power Blurring \cite{power-blurring} is another temperature calculating model, which is developed based on a matrix convolution approach to reduce computation time as compared to the finite difference method. Power blurring (PB) uses a technique analogous to image blurring for calculating temperature distributions. Sarangi et al. \cite{sarangi-thermal-survey} presents one of the most comprehensive and updated surveys of thermal estimation and modeling tools. The semiconductor industry has also developed several comprehensive thermal modeling and estimation tools. While many of these tend to be proprietary, some like Intel Docea \cite{intel-docea} tool is available for experimental evaluation. Thermal simulation algorithms for calculating the on-chip temperature distribution in a multilayered substrate structure rely on Green's function and discrete cosine transforms (DCT). Varshney et al. \cite{nanotherm} present NanoTherm, a solution to compute Green's function using a fast analytical approach that exploits the symmetry in the thermal distribution. Additionally, conventional methods fail to hold at the nanometer level, where it is necessary to solve the Boltzmann transport equation (BTE) to account for quantum mechanical effects, without which, there can be errors in temperature calculation of upto 60\%. NanoTherm also provides a fast analytical approach to solve the BTE for nanometer chip designs.
	

\subsection{Accelerator Simulators}
\label{subsection:accelerator-sim}
With the rise of domain-specific accelerators, the need for power and performance modeling of such chips has become an important area of research. Accelerators could be GPUs, application specific integrated circuits (ASICs), digital signal processors (DSP), field programmable gate arrays (FPGA), near-data and in-memory processing engine, or any other similar component optimized for fixed functions. 
	
	
Alladin \cite{aladdin-accel-sim} is a pre-RTL power and performance modeling framework for accelerators. The framework takes high-level language descriptions of algorithms as inputs, and uses dynamic data dependence graphs (DDDG) as a representation of an accelerator without having to generate RTL. Starting with an unconstrained program DDDG, which corresponds to an initial representation of accelerator hardware, Aladdin applies optimizations as well as constraints to the graph to create a realistic model of accelerator activity and then overlays power and performance estimation. To accurately model the power of accelerators, Aladdin uses precise activity factors, accurate power characterization of different DDDG components, characterizes switching, internal, and leakage power from design compilers for each type of DDDG node (multipliers, adders, shifters) and registers. Minerva \cite{minerva-dnn-sim} is a highly automated co-design approach across the algorithm, architecture, and circuit levels to optimize DNN hardware accelerators. It allows for the modeling and simulation of ultra-low power DNN accelerators (in the range of tens of milliwatts), making it feasible to deploy DNNs in power-constrained IoT and mobile devices. More accelerator simulators are described in detail in Akram et al. \cite{ayaz-lina-2016}.

\section{System Level Techniques for Energy Efficiency}
\label{section:system-level-techniques}
In this section we look at how underlying architectural and microarchitectural techniques are used at higher levels of the software hierarchy (firmware, operating system and applications) and how energy efficiency is implemented at the entire system. Depending on the constraints of the system (IoT, wearable, smartphone, or server) several of these techniques may be used to fine tune the system for specific workloads. Since it is hard to discuss system level techniques without being specific about the underlying system architecture, we elaborate on ARM and x86 systems. We will first cover the system level techniques implemented in these systems and then discuss how software uses these features to optimize for energy efficiency.

\subsection{ARM System Architecture and Energy Efficiency Features}
\label{subsection:arm-pm}
	
	
\subsubsection{Clock Gating, Dormant Mode and Power Collapse}
\label{subsubsec:arm-pm-cg-power-collapse}
ARM processors implement clock gating for the CPU using the Wait-For-Idle (WFI) instruction. Most ARM cores also provide the capability to clock gate the L2 cache, debug logic, and other components using co-processor instructions. Dormant Mode allows for cache controller and CPU to be powered down with the cache memories remaining powered on. The cached RAMs may be held in a low-power retention state where they keep their contents but are not otherwise functional. This mode helps achieve power savings by turning off the cache masters at the same time preventing any performance hit due to invalidation/flush of the caches. Power gating a core results in the context having to be reset at resume. ARM based platforms may have multiple clusters of cores, with each cluster having a shared L2. Power collapse of all CPU cores in a cluster results in a cluster power down which includes disabling cache snoops and power gating the L2 cache. A System Control Processor (SCP) provides several PM functions and services – (a) Managing clocks, voltage regulators to support DVFS (b) Power state management for SoC domains and (c) Maintain/enforce consistency between device states within the system. 
	

\subsubsection{DVS/DVFS/AVFS}
All modern ARM SoCs usually support software controlled DVFS. Apart from a maximum sustained frequency, several ARM SoC vendors add a boost mode where the CPU can be overclocked if required. For Symmetric Multi Processors (SMP) and Hetergeneous Multi Processor (HMP) systems with multiple (hetero) cores, the most common configuration is having a single voltage rail for all the cores in a cluster. Per-core voltage rail implementations are rare due to design complexity. Per-core clock lines are available on some SoCs allowing for independent control of core frequency with glue logic handling the voltage synchronization for the common voltage rail. ARM11 introduced a new \textit{Intelligent Energy Manager (IEM)} that could dynamically predict the lowest voltage. This is \textit{Adaptive Voltage Frequency Scaling (AVFS)} - a closed-loop system which continuously monitors system parameters through sensors. The IEM lowers the voltages below the values of the stock voltage tables when silicon characteristics reported by sensors permit it. Some ARM-based SOCs use power-efficient and high performance hetero cores in a single SoC as separate clusters, called \textit{BIG.LITTLE} systems. The standard pattern of usage on mobile devices is that of periods of high processing and longer periods of light load. The core idea is that with appropriate task placement and packing on the HMP clusters, performance and power criteria both can be met. The recent DynamIQ is similar - it bundles both high performance big CPUs and high efficiency LITTLE CPUs into a single cluster with a shared coherent memory. All task migrations between big and LITTLE CPUs take place within a single CPU cluster through a shared memory, with the help of an upgraded snoop management system, resulting in improved energy efficiency. The transfer of shared data between BIG and LITTLE cores takes place within the cluster reducing the amount of traffic being generated and in turn the amount of power spent.
	
\subsubsection{Device PM and Power Domains}
\label{subsubsec:arm-device-pm}
ARM SoCs are typically partitioned into multiple voltage domains allowing for independent power control of devices and independent DVFS. Additionally voltage regulators are organized hierarchically so that the Linux Regulator framework can be used by software to indicate when components are idle and do not need clock/power. This allows for system level power collapse. Power collapse of an IP or group of IPs is made possible by this partitioning and hierarchical clock and voltage framework. The focus is always to reduce the number of always-on power domains on a platform and allow as many domains as possible to be turned off. Software orchestrates these dynamic power plane management based on the usage scenario - device drivers manage the clock and power to respective hardware and OS software manages system level power domains. The common system low power states on ARM SoCs are:
	\begin{itemize}
		\item \textbf{S2R}: Here the entire system is off except for components like wake-up logic and internal SRAMs
		\item \textbf{Low Power Audio}: Most SoCs support a special low power audio state to minimize power consumption for use cases like “screen off user listening to music”. The internal audio SRAM, DRAM, DMA and I2S Controller are only active (audio power domain is ON). CPU/dedicated DSP wakes up periodically to process the audio data and the display remains off.
		\item \textbf{Low Power Display}: Another common use case is when the modem, display and audio are only active during a voice call. This is handled by a low power display state.
	\end{itemize}
    Several other similar low power states are supported based on the low power usage scenario (low power sensing, low power voice call). Suspend-to-Disk, which is a common feature in larger laptops and desktops, is generally not supported on ARM based tablets/mobiles due to large resume latencies.
	
	
	
	\subsection{Intel x86 Power Management}
	\label{subsection:intel-pm}
	Intel x86 SOCs provide fine-grained knobs for device and system level power management. OS Power managers like ACPI traditionally directs the platform to various power states (S3/S4, for example) depending on different power policy set by the user. Intel SOCs have components in OS and firmware that guide the power states for the CPU, devices, other subsystems and the system as a whole. A combination of hardware (dedicated power management units) and software (OS, kernel drivers, software) orchestrate the transition of the system into low power states. The overall power management architecture is built around the idea of aggressively turning off subsystems without affecting the end user functionality and usability of the system. This is enabled by several platform hardware and software changes:
	\begin{itemize}
		\item \textit{On die clock/power gating} - applicable to all subsystems, controllers, fabrics and peripherals.
		\item \textit{CPU C-states} - C-states are the CPU cores' low power states and a state Cx, means one or more subsystems of the CPU is at idle, powered down. For example, C1 is a AUTOHALT state, C3 means that the processor caches are flushed and the processor clocks are shutoff. In C6, the CPU core voltage can be shut off. More details are in the Intel x86 developer manual \cite{intel-x86-dev-manual}. Higher levels of software (operating system) can initiate entry into some of these states and monitor residencies in different states.
		\item \textit{CPU P-states} - CPU P-states are performance states, and each Px state represents a specific operating frequency and a corresponding voltage it needs to run at. More details are in the Intel x86 developer manual \cite{intel-x86-dev-manual}. Selecting an appropriate P-state can be done through architectural registers, and there are several software and hardware-software techniques to do this, which we will describe shortly.
		\item \textit{Subsystem active idle states} – applicable to all OS/driver controlled components. These states, called \textbf{D0ix}, are managed either in hardware or using the Linux Runtime PM framework (in the kernel) and the device drivers (in the OS). 
		\item \textit{Platform idle states} - extending idleness to the entire platform when all devices are idle. These are termed \textbf{S0ix} states. In these states, many platform components are transitioned to an appropriate lower power state (CPU in low power sleep state, memory in self refresh, and most components are clock or power gated).
		\item Microcontrollers for power management of north (CPU, GPU) and south complex IPs (peripherals) respectively. The microcontrollers coordinate device and system transitions, voltage rail management, and system wake processing.
		\item \textit{Integrated Voltage Regulators (IVR)}: On-die and on-chip voltage regulators provide fine-grained power delivery to different parts of the chip and this is managed by hardware and/or firmware/software.
	\end{itemize}
	


	Many Intel SoCs have CPU cores organized in a hierarchical structure, which has three levels: core, module, and package. A package contains two modules, each of which groups two cores together.  This topology allows two levels of task consolidation: in-package and in-module. With in-package consolidation, the workload runs on either the first module or both modules, i.e., all of the four cores. Intel CPUs support DVFS or performance states (or P-states) for OS controlled management of processor performance. The P-states are exposed via ACPI tables to the OS. OS Software requests a P-State based on performance needs of the application (in Linux/Android, this is via the cpufreq-based governors). Atom cores also support Turbo frequencies akin to boost on ARM SoCs. Turbo allows processor cores to run faster than the “guaranteed” operating frequency if the processor is operating below rated power, temperature, and current specification limits of the system. Turbo takes advantage of the fact that the rated maximum operating point of a processor is based on fairly conservative conditions which occur infrequently. 


\subsubsection{System low power states}
Intel SoCs support the following transient low power system states:
	\begin{enumerate}
		\item S0ix: Shallow idle state for the entire SOC
		\item S0ix-Display: display can be kept in a shallow low power state, with display controller periodically waking up to feed the contents of the display panel and the rest of the SOC fully powered off.
		\item S0ix-Audio: SOC in low power state except audio block.
		\item S0ix-Sensing: SOC in low power state except sensor hub to support several low power sensing modes such as pedometer
		\item S0i3: Entire SOC is in low power state, except for wake logic/sequencing and a small amount of memory to store code for restoring the SOC back to operating state.
	\end{enumerate}

All these states are transparent to applications and are entered/exited by close orchestration between operating system, firmware, microcontrollers and hardware and have different entry/exit latencies. In addition to these, systems generally support \textbf{Suspend-to-RAM}, where the entire system is off except for minor exceptions such as wake-up logic, internal SRAMs etc. and \textbf{Suspend-to-Disk}, that has larger entry/exit latencies but also deeper power savings. 

\subsection{OS and Software Techniques}
\label{subsection:os-sw-pm}
	Linux \cite{linux-kernel-doc} has developed several energy efficiency features in the last two decades and the following have been among the most important ones:
	
	\begin{enumerate}
	    \item \textbf{Timers and Tickless Scheduling}: 	The scheduler allocates CPU time to individual processes via interrupts. Programmable timer interrupts keep track of, and handle future events. In traditional systems we had a periodic tick i.e. the scheduler runs at a constant frequency. This resulted in periodic wake-ups and poor energy efficiency. Linux evolved to use three primary mechanisms, as described in Siddha et al. \cite{linux/tickless} and \cite{linux/timers} - (a) \textit{Dynamic tick} - program the next timer interrupt to happen only when work needs to be done, (b) \textit{Deferrable timers} - bundle unimportant timer events with the next interrupt (c) \textit{Timer migration} - move timer events away from idle CPUs. Some CPUs also support \textit{power-aware interrupt redirection (PAIR)}, that ensures that interrupts are directed to already-awake CPU cores, rather than wake up a sleeping core.
	    
	    \item \textbf{CPUFreq}: This is a standard Linux framework used for CPU Dynamic Voltage and Frequency Scaling (DVFS). Processors have a range of frequencies and corresponding voltages over which they may operate. The CPUFreq framework allows for control of these voltage-frequency pairs according to the load through components called \textit{governors}. There are several different governors based on how the algorithm can be controlled and implemented. The performance governor is used for optimizing CPU performance whereas the power-save governor aims to conserve energy. The user-mode governor allows a user space application to control the DVFS states. The on-demand governor was one of the most popular governors, described in Pallipadi et al. \cite{linux/on-demand}. More recently, the interactive governor was developed for mobile devices that require optimized burst performance for on-screen usages. The Intel P-state driver is slightly different - it can operate in two different modes, active or passive. In the active mode, it uses its own internal performance scaling governor algorithm or allows the hardware to do performance scaling by itself, while in the passive mode it responds to requests made by a generic CPUFreq governor implementing a certain performance scaling algorithm. All of these are described in detail in the Linux kernel documentation \cite{linux/dvfs} and the Intel P-state driver is described in more detail in \cite{intel/p-state-driver}. 
	    
	    \item \textbf{CPU Idle}: This is a Linux kernel subsystem that manages the CPU when it is idle and the core idea is to \textit{do nothing, efficiently} (Pallipadi et al. \cite{pallipadi07cpuidle}). Usually, several idle states, known as C-states, are supported by the processor. The convention for C-state naming is that 0 is active state and a higher number indicates a deeper idle state e.g. C1-Clock Gating. Deeper idle states mean larger power savings as well as longer entry/exit latencies. The inputs required by the framework for C-state entry are –  CPU idleness, next expected event, latency constraints, break-even time and exit latency. Based on the inputs, a specific C-state is entered via architecture specific instructions such as MWAIT in x86.
	    
	    \item \textbf{PM Quality of Service}: PM QOS is a latency and performance control framework in Linux \cite{linux/pmqos}. It provides a synchronization mechanism across power managed resources with a minimum performance need as expressed by a device. The kernel infrastructure facilitates the communication of latency and throughput needs among devices, system, and users. QoS can be used to guarantee a minimum CPU frequency level to meet video playback performance or to limit the max device frequency to reduce skin temperature, and similar constraints.
	    
	    \item \textbf{Voltage Regulator framework} is a standard kernel interface to control voltage/current regulators \cite{linux/vrframework}. It is mostly used to enable/disable a regulator output or control the output voltage and or current. The intention is to allow systems to dynamically control regulator power output in order to save power and prolong battery life. This applies to both voltage regulators (where voltage output is controllable) and current sinks (where current limit is controllable). Many drivers use this framework to enable/disable voltage rails or control the output of low drop out oscillators (LDOs) or buck boost regulators.
	    
	    \item \textbf{Runtime PM framework} is a widely used framework in the Linux kernel \cite{linux/runtimepm} to reduce the individual device power consumption when the device is idle through clock gating, gating the interface clock, power gating or turning off the voltage rail. In each of the cases we need to ensure that before we move the device to a low power state, any dependent devices are also considered. The framework allows for understanding and defining this tree for hierarchical control. 
	    
	    \item \textbf{Devfreq} framework is used for handling DVFS of non-CPU devices such as GPU, memory and accelerator subsystems \cite{linux/devfreq}. Devfreq is similar to cpufreq but cpufreq does not allow multiple device registration and is not suitable for heterogeneous devices with different governors. It exposes controls for adjusting frequency through sysfs files which are similar to the cpufreq subsystem. 
	    
	    \item \textbf{System sleep states} provide significant power savings by putting much of the hardware into low power modes. The sleep states supported by the Linux kernel are power-on standby, suspend-to-RAM (S2R), suspend to idle (S2I) and suspend to disk (hibernate) \cite{linux/sys-states}. Suspend to idle is purely software driven and involves keeping the CPUs in their deepest idle state as much as possible. Power-on standby involves placing devices in low power states and powering off all non-boot CPUs. Suspend to RAM goes further by powering off all CPUs and putting the memory into self-refresh. Lastly, suspend to disk gets the greatest power savings through powering off as much of the system as possible, including the memory. The contents of memory are written to disk at suspend, and on resume this is read back into memory.
	    
	    \item \textbf{Power Capping Framework}: The Linux power capping framework provides a consistent interface between the kernel and the user space that allows power capping drivers to expose the settings to user space in a uniform way \cite{linux/powercap}. Power zones represent different parts of the system, which can be controlled and monitored using the power capping method determined by the control type the given zone belongs to. They each contain attributes for monitoring power, as well as controls represented in the form of power constraints. With the power capping framework, it is possible to apply power capping to a set of devices together. Intel RAPL [Section \ref{subsubsection:intel-rapl}] is one form of a power capping framework.
	    
	    \item \textbf{Multi-cluster PM and Energy Aware Scheduler}: The Multi Cluster PM (MCPM) layer supports power modes for multiple clusters. It implements powering up/down transitions of clusters including the necessary synchronization. The Linux scheduler traditionally placed importance on CPU performance and did not consider the different power curves if disparate cores exists in one system. The Energy Aware Scheduler (EAS) links several otherwise independent frameworks such as CPUFreq, CPUIdle, thermal and scheduler to be more energy efficient even for disparate cores. A scheduler directed CPUFreq governor called schedutil has been introduced which takes optimal decisions regarding task placements, CPU idling, frequency level to run, among other parameters. Based on a SoC specific energy model, EAS realizes a power efficient system with minimal performance impact. This is commonly implemented today on several ARM based systems \cite{arm/eas}. 
	\end{enumerate}

	\subsection{System and OS Techniques for Energy Efficiency in GPUs}
    \label{subsection:system-tech-gpus}
    The techniques for improving energy efficiency of GPUs overlaps with those used for CPUs and a detailed survey is presented in Mittal et al. \cite{mittal-vetter-ee-gpu}. Some key techniques are highlighted here:
	\begin{enumerate}
	    \item \textbf{Workload-based dynamic resource allocation}: This is based on the observation that the power consumption of GPUs is primarily dependent on the ratio of global memory transactions to computation instructions and the rate of issuing instructions. The two metrics decide whether an application is memory intensive or computation intensive respectively. Based on the metrics, the frequency of GPU cores and memory is adjusted to save energy. Some systems use an integrated power and performance prediction system to save energy in GPUs. For a given GPU kernel, their method predicts both performance and power and then uses these predictions to choose the optimal number of cores that can lead to the highest performance per watt value. Based on this, only the desired number of cores can be activated, while the remaining cores can be turned off using power gating.
	    \item \textbf{CPU-GPU Work division}: Research has shown that different ratios of work division between CPUs and GPUs may lead to different performance and energy efficiency levels. Based on this observation, several techniques have been implemented that dynamically choose between CPU and GPU as a platform of execution of a kernel based on the expected energy efficiency on those platforms. 
	    \item \textbf{CPU-GPU Power Sharing}: In several recent CPU-GPU systems, dynamic power sharing is implemented at the firmware, microkernel and/or OS level to dynamically balance the power being consumed by the CPUs and GPUs. For example, in \cite{intel-cpu-gpu-power-sharing}, the  power sharing framework is used to balance the power between high performing processors and graphics subsystem. It helps to manage temperature, power delivery and performance state in real time and allows system designers to adjust the ratio of power sharing between the processor and graphics based on workloads and usages. 
	 \end{enumerate}
    
\section{Recent advances in SOC and System Level Energy Efficiency}
\label{subsection:recent-system-level-pm}
The last few years have seen rapid innovations in SOC design/microarchitecture and system level power/performance optimizations across x86 and non-x86 architectures, including the rise of custom-designed ARM chips by different companies such as Apple, Amazon, Google, Ampere, etc. In this section, we review some of the key technologies across these architectures and systems. 

\subsection{Energy Efficiency in Intel Processors/SOCs}
\label{subsection:intel-x86-pm}
Starting with the Broadwell, Intel architectures implemented several system level techniques for energy efficiency while pushing the performance envelopes for newer workloads and all-day battery life for active scenarios. Similarly, AMD processors also evolved several techniques across client and server processors. Some of the most important features and techniques are described here and are summarized in Table \ref{tab:intel-amd-ee}. 

   \begin{table}
    	\caption{Summary of recent energy efficiency techniques in Intel and AMD x86 processors}
		\label{tab:intel-amd-ee}
    \begin{tabular}{|c|l|}
        \toprule
        \textbf{Technique} & \textbf{Processor/SOC family}\\
        \midrule
        \multicolumn{1}{|m{6cm}}{Per Core P-States, Uncore Frequency Scaling, Integrated Voltage Regulator (IVR), Running Average Power Limiting (RAPL)} & 
		\multicolumn{1}{|m{6cm}|}{Intel Haswell (22nm) \cite{haswell-ee}} \\ \hline
        \multicolumn{1}{|m{6cm}}{Intel Speedshift (Hardware P-states), Energy-aware Race to Halt (EARtH), Energy Performance Bias (EPB) / Energy Performance Preference (EPP), Hardware/SOC Duty Cycle, Memory DVFS, enhanced RAPL, IccMax/Peak current management} & 
		\multicolumn{1}{|m{6cm}|}{Intel Skylake (14nm) \cite{skylake-arch-ee}} \\ \hline
        \multicolumn{1}{|m{6cm}}{Dynamic Tuning (ML-based Turbo)} & 
		\multicolumn{1}{|m{6cm}|}{Intel Ice Lake (10nm) \cite{icelake-arch-ee}} \\ \hline
        \multicolumn{1}{|m{6cm}}{Autonomous Fabric and Memory DVFS, independent clock and power domains for Graphics, Memory, PCIe, USB, Thunderbolt} & 
		\multicolumn{1}{|m{6cm}|}{Intel Tiger Lake (10nm) \cite{tigerlake-arch-ee}} \\ \hline
        \multicolumn{1}{|m{6cm}}{Heterogeneous x86 cores} & 
		\multicolumn{1}{|m{6cm}|}{Intel Lakefield \cite{intel-lakefield}} \\ \hline
        \multicolumn{1}{|m{6cm}}{Locally Efficient Application Power Management (LEAPM), Globally Efficient APM (GEAPM), Core Bound Boost (CBB), Memory-Bound Boost (MBB)} & 
		\multicolumn{1}{|m{6cm}|}{AMD (28nm) \cite{amd-soc-pm}} \\ \hline
        \bottomrule
    \end{tabular}
    \end{table}

\subsubsection{Intel Speed Shift Technology (Hardware P-States)}
\label{subsubsection:intel-hwp}
Skylake (Doweck et al. \cite{skylake-arch-ee}) is a SOC consisting of 2-4 CPU cores, Graphics, media, a ring interconnect, an integrated system system, and a Power Control Unit (PCU) that houses the power management firmware logic and provides interfaces to higher power management hierarchies (BIOS, OS, device drivers, etc.). Speed Shift is a faster response vehicle to frequency requests and race to sleep by migrating the control from the operating system back down to the hardware. Current implementation of OS-guided P-states can take up to 30 milliseconds to adjust, whereas if they are managed by the processor, it can be reduced to about 1 millisecond. At any time the OS can demand control of the states back from the hardware if specific performance is needed. The key concept behind autonomous processor level control is to find the power state that uses the least total system system power, and stay in that state as often as possible.

\subsubsection{Workload Aware Power Balancer}
\label{subsubsection:intel-power-balancer}
For active workloads, Intel Skylake looks to balance the power across CPU cores, Graphics, and other subsystems (memory and uncore). A feedback-based control system monitors the different units (CPU, Graphics, memory, uncore, imaging/camera subsystem) and uses that information to understand the nature of the workload. That information is used to split the available system power between CPU, Graphics and other subsystems. 
By default, such power budget allocation could be fixed, corresponding to worst-case performance demands/workloads, even if the domains are under utilized. This unfair allocation is sub-optimal and can hamper overall system performance and throughput. In SysScale \cite{intel/skylake/sysscale}, the authors introduce an algorithm to predict the performance demands (bandwidth, latency) of the SOC domains and implements a new DVFS algorithm to distribute SOC power based on predicted performance demands. Furthermore, in addition to a global DVFS mechanism, SysScale optimizes the DVFS of each domain from an energy efficiency perspective. 

\subsubsection{SOC/ Hardware Duty Cycling}
\label{subsubsection:intel-skylake-hdc}
One of the fundamental concepts for saving power is \textit{race to idle or sleep} - get the job done as soon as possible, and put the CPU into an idle state. As process technology starts encountering fundamental physics limits, and due to the fact that transistors cannot operate reliably below a certain threshold voltage, idling the processor at lower frequencies starts providing diminishing returns once we get closer to the threshold voltage. Intel Broadwell and Haswell processors introduced the idea of Duty Cycling Control (DCC) for the integrated graphics unit, which meant that the GPU would be cycling between on and off states. Skylake introduced this concept for the CPU cores as well, and rapidly transitions the CPU cores between on and off states. This technique has shown to save large amounts of power for a range of workloads.

\subsubsection{Energy Aware Race to Halt}
Intel Skylake also introduced a new algorithm called Energy Aware Race to Halt (EARtH) as described in Deweck et al. \cite{intel-skylake-earth}. The motivation behind this is based on the observation that controlling CPU power has limited impact on the overall energy efficiency of the computing platform due to energy consumption of other platform components. When the CPU power dominates total power, the minimum energy is achieved when the CPU operates at the lowest frequency mode (LFM). When the rest of the platform consumes significantly higher power than the CPU, the most energy efficient policy is Race To Halt (RtH). In many real systems, however, power is balanced between CPU and the rest of the platform for different workloads. In such systems the minimum energy point may happen at some intermediate frequency. The authors in this paper demonstrate this observation in real systems and with real production workloads and present an Energy Aware Race to Halt (EARtH) algorithm that identifies that minimum energy point at run time. Starting with Skylake, this algorithm (with some enhancements) is now available in most Intel Core processors including the latest Ice Lake and Tiger Lake SOCs. 

\subsubsection{Running Average Power Limiter (RAPL)}
\label{subsubsection:intel-rapl}
Intel's RAPL provides a set of counters providing energy and power consumption information using a software power model that estimates energy usage by using hardware performance counters and I/O models \cite{intel/rapl}. The key idea behind RAPL is that of Thermal Design Power (TDP). The TDP of a system represents the maximum amount of power the cooling system in a computer is required to dissipate. For example, for a processor with TDP of 35W, Intel guarantees the OEM that if it implements a chassis and cooling system capable of dissipating that much heat, the chip will operate as intended. This is the power budget under which the system needs to operate. But this is not the same as the maximum power the processor can consume. It is possible for the processor to consume more than the TDP power for a short period of time without it being “thermally significant”. Using basic physics, heat will take some time to propagate, so a short burst may not necessarily violate TDP. RAPL provides a set of counters providing energy and power consumption information. RAPL is not an analog power meter, but rather uses a software power model. This software power model estimates energy usage by using hardware performance counters and I/O models. 

RAPL provides a way to set power limits on processor packages and DRAM. This will allow a monitoring and control program to dynamically limit max average power, to match its expected power and cooling budget. In addition, power limits in a rack enable power budgeting across the rack distribution. By dynamically monitoring the feedback of power consumption, power limits can be reassigned based on use and workloads. Because multiple bursts of heavy workloads will eventually cause the ambient temperature to rise, reducing the rate of heat transfer, one uniform power limit can’t be enforced. RAPL provides a way to set short term and longer term averaging windows for power limits. These window sizes and power limits can be adjusted dynamically.

\subsubsection{Intel P-state driver}
\label{subsubsection:intel-p-state}
Starting with Intel's Sandybridge, this driver provides an interface to control the P-State selection for the processors. The underlying driver in the kernel is essentially a Proportional Integral Derivative (PID) controller with software-tunable interfaces to control each of the P, I, and D parameters \cite{intel/p-state-driver}. The driver decides what P-State to use based on the requested policy from the OS's cpufreq core. If the processor is capable of selecting its next P-State internally, then the driver will offload this responsibility to the processor (Hardware P-States). If not, the driver implements algorithms to select the next P-State. The P-state driver is primarily supported only for Linux based platforms. 

\subsubsection{Dynamic Current Management, peak current and thermal protection}
\label{subsubsection:intel-iccmax}
Intel processors have two modes of current and thermal protection: throttling,and automatic shutdown. As described in \cite{intel-peak-current-thermal} and \cite{intel-skylake-ee}, when a core exceeds the set throttle temperature, it will start to reduce power to bring the temperature back below that point. The throttle temperature can vary by processor and BIOS settings, and the throttling actions can be different (turning down display, turning off charging for example). Peak current violations are handled similarly. If the conditions are such that throttling is unable to keep the temperature down, such as a thermal solution failure or incorrect assembly, the processor will automatically shut down to prevent permanent damage. The peak current and peak thermal limits respectively are controlled by Processor Core IccMax and Thermal Limit PL1/PL2/PL3 settings in the BIOS. 

\subsubsection{Connected Standby}
\label{subsubsection:connected-standby}
Connected Standby is a feature used in laptops, tablets, and smartphones in order to reduce energy consumption when the device is fully idle, while remaining connected to communication channels. A mobile device enters the deepest-runtime-idle-power state (DRIPS), which minimizes power consumption and retains fast wake-up capability. Haj-Yahya et al. \cite{intel-skylake-odrips} look at ways to increase battery life in the connected-standby mode and implement an optimized DRIPS (ODRIPS) mechanism. ODRIPS is based on 2 key ideas: (1) offload wake event monitoring to low-power off-chip circuitry, which enables turning off most of the SOC 2) offload processor context to off-chip storage (DRAM), thus eliminating the need for on-chip high-leakage SRAMs and thereby reducing leakage power. 

\subsubsection{Thermal Management - Intel DPTF and ARM Intelligent Power Allocator}
\label{subsubsection:thermal-mgmt}
Smart system level thermal management has improved over the last decade as form factors and workloads have impacted platform thermals significantly. Platforms today encompass several thermal sensors - per-CPU, per-GPU, for the connectivity radios, USB subsystem, etc. An intelligent thermal manager needs to comprehend the data from thermal sensors, estimate possible platform level impact (skin temperature of a device needs to be calculated using different equations based on the individual thermal sensor readings), and then impose policies (such as throttling the CPU, dim the display, or disable charging) to ensure the system can continue to function. 

The Intel Dynamic Platform and Thermal Framework (Intel DPTF) is implemented on both Linux and Windows platforms \cite{intel/dptf}. It includes the DPTF Framework Manager, Policies, and Participants. The DPTF manager is responsible for all communication into the user space code, and serves as the interface to Eco-System Independent Framework (ESIF). It manages events and notifications to/from the ESIF layer and is responsible for high level arbitration of policies to ensure system level thermal management. DPTF policies are the intelligent plug-in glue that determines what needs to be done to address specific thermal situations. DPTF participants are the entities that expose telemetry (CPU temperature, for example) and provide controls (throttling the CPU P-states). 

Similarly, ARM's Intelligent Power Allocator (IPA) \cite{arm/ipa} performs proactive power and thermal management by continuously adapting response based on power consumption and thermal headroom. It implements a closed-loop Proportional Integral Derivative (PID) controller for accurate temperature control. The Dynamic Power Partitioning component optimally allocates power to CPU and GPU based on the current workload based on a SoC power model, which is composed of voltage/frequency operating point for each key IP block (e.g. CPU, GPU). IPA thus  maps between runtime power consumption (measured through on-chip monitoring counters) and theoretical operating points of each component.

\subsection{Energy Efficiency in AMD Processors/SOCs}
\label{subsection:amd-ee}
AMD processors and SOCs support several energy efficiency features including clock and power gating, DFS, DVS, DVFS, link level power management, etc. Some of the recent advances are noted here. 

AMD SOC's power management is described in detail in Bircher et al. \cite{amd-soc-pm}. Here the authors describe several aspects of AMD's SOC/system level power management. The power manager is implemented on an on-die microcontroller that uses power and thermal feedback from the SOC through digital power monitors. The power monitor accounts for fluctuations in dynamic power caused by the workload and also accounts for the effects of voltage, frequency and temperature using built-in models. To provide consistent repeatable performance, the models are calibrated for each version/model of the SOC. The power manager contains three performance controllers: Global Efficient Application Power Management (GEAPM), Core-Bound Boost (CBB) and Memory-Bound Boost (MBB). Another feature, called Locally Efficient APM (LEAPM) is also implemented for IP-level power management. 
\begin{enumerate}
    \item GEAPM optimizes the balance of power between CPU and GPU within an SOC. It is global in the sense that it seeks to maximize the SOC-level performance rather than the individual (local) performance of either CPU or GPU. 
    \item The CBB and MBB features improve performance of CPU-centric workloads. CBB increases CPU performance by shifting power from the memory subsystem to the CPU for core-bound workloads (with little memory dependence). 
    \item The MBB feature detects memory latency-sensitive workloads and shifts power to the memory controller. 
    \item LEAPM works on the premise that some power management decisions can be made using only information local to the IP and it essentially uses different ways of tracking IP-level utilization to determine when to shift power away from an IP.
\end{enumerate}
All of these features shift power from other parts of the system that have less impact on performance to those with more power requirements thus providing higher performance in a constrained environment.
AMD also optimizes power consumption for different workloads through different processor/SOC settings based on die temperature, expected leakage (as leakage depends on temperature), part-to-part variations in the die itself, as documented in Suggs et al. \cite{amd/micro/zen2}, Arora et al. \cite{amd/hotchips/AroraBW20} and \cite{amd/epyc/workload-tuning}. 

\subsection{The rise of ARM in enterprise, HPC and the Cloud}
\label{subsection:arm-hpc-datacenter}
The last couple of years has also seen the rise of ARM architectures in data center and cloud systems that were traditionally x86-based, which was primarily due to the unmatched performance of Intel and AMD processors. 


Amazon Web Services have released custom-build high performance ARM processors for the cloud, named Graviton, Graviton 2, and more recently, Graviton 3. These are available as AWS's EC2 instances. Graviton is an ARM64 processor based on A72 microarchietcture. In \cite{aws/arm64}, the authors perform detailed performance analysis of AWS's Graviton A1 against similar class of Intel Xeon processors and observe that the A1 achieves almost similar performance in web services, with significant cost savings across various video and database workloads. Graviton2 improves on this and delivers enhanced price-performance by 40\% in comparison to present generation x86-fueled processors. At the time of this writing, Graviton 3 is touted to be 25 percent faster than Graviton 2, with 2x faster floating-point performances, and a 3x speedup for machine learning workloads, with a 60X energy reduction \cite{aws/graviton3}.


With its highly improved power/performance/cost benefits, ARM architectures and processors have also made a big headway to HPC and supercomputing systems \cite{hpc/arm}. 

\section{Verification}
\label{section:verification}
    Verifying energy efficiency features of complex SOCs is a big challenge from hardware as well as a system level perspective, since power management flows span the entire platform. Ideally, each system component (hardware, firmware, software) needs to be verified for its power management capability both individually as well as how they work in relation to other components, and with real workloads. In addition, system-level power flows (low power idle/standby states) also need to be verified before silicon tape-in is achieved. Power management brings a host of new types of bugs which are not in the class of traditional functional bugs. Table \ref{tab:pm-bugs} shows the different classes of bugs and the new verification techniques required, some of which are hard to verify in pre-silicon (for example, voltage sequencing, due to lack of integrated power delivery models into SOC emulation models) or thermal runways (these are usually verified on form factor devices in thermal chambers that simulate different thermal conditions and heat flows). At a high level, verification can be done at either the gate level, RTL/architectural level or at SOC/system level.
	
    \begin{table}
    	\caption{Summary of Power Related Bugs}
		\label{tab:pm-bugs}
    \begin{tabular}{|c|l|}
        \toprule
        \textbf{Power Related Issue} & \textbf{Verification techniques required} \\
        \midrule
        \multicolumn{1}{|m{6cm}}{Isolation/level shifting bugs} & 
		\multicolumn{1}{|m{6cm}|}{Verify connection, placement, isolation/level shifting} \\ \hline
		\multicolumn{1}{|m{6cm}}{Control sequencing bugs} & 
		\multicolumn{1}{|m{6cm}|}{Include power intent files like UPF}\\ \hline
		\multicolumn{1}{|m{6cm}}{Electrical problems like memory corruption} & 
		\multicolumn{1}{|m{6cm}|}{Reach good power state coverage} \\ \hline
			
		\multicolumn{1}{|m{6cm}}{Power/voltage sequencing bugs} & 
		\multicolumn{1}{|m{6cm}|}{Verify FW/SW control sequences}\\ \hline
			
		\multicolumn{1}{|m{6cm}}{Power gating collapse/dysfunction, Clock domain/crossover bugs} & 
		\multicolumn{1}{|m{6cm}|}{Verification at each stage of design, not just RTL; verify netlist at each handoff, power switch/rail connectivity}\\ \hline
			
		\multicolumn{1}{|m{6cm}}{Power-on/reset bugs} & 
		\multicolumn{1}{|m{6cm}|}{Wide coverage of test cases across power-on/reset flows}\\ \hline
			
		\multicolumn{1}{|m{6cm}}{Thermal runways/cooling inefficiencies} & 
		\multicolumn{1}{|m{6cm}|}{Verify thermal conditions, thermal modeling for different form factors/designs} \\ \hline
		
		\multicolumn{1}{|m{6cm}}{Bugs due to concurrent access from multiple IPs during end-to-end use cases} & \multicolumn{1}{|m{6cm}|}{Verify end to end system level power sequences, including FW, SW, drivers to uncover race conditions} \\ \hline
        \bottomrule
    \end{tabular}
    \end{table}

\subsection{Verification of Low Power transformations at gate level}
 \label{subsection:gate-level-verification}
 \textit{Formal verification}, especially equivalence checking, has achieved considerable success in the context of low power verification. \textit{Combinational equivalence checking} checks two acyclic, gate-level circuits. Combinational equivalence checkers can also be used to check equivalence of two sequential designs, provided the state encodings of the two designs are the same. Although this technique has widespread use in many commercial tools, the real challenge of sequential verification is in verifying two designs with different state encodings. Sequential satisfiability engines, like the one in Lu et al.  \cite{verification-seq-sat-solver} and sequential ATPG engines (Abraham et al.  \cite{jabraham-seq-atpg-verification}) solves this problem to a large extent by unrolling the circuit until a given time frame. However, these techniques operate at the gate level, where they reason in the Boolean domain.
 
\subsection{Verification of Low Power transformations at RTL/architectural level}
\label{subsection:rtl-level-verification}
Given the nature of power management and the hardness of the problem at lower levels of design, more verification is usually focused on RTL and higher levels of abstraction. In Silveira et al. \cite{verification-power-gate-rtl}, the authors describe methods to verify RTL power gating through transaction level models. Some attempts have been made to apply sequential equivalence checking to the behavioral RTL descriptions of designs. Semeria et al. \cite{rtl-c-verification} describe a methodology for checking the combinational equivalence between C and RTL is described. In Viswanath et al. (\cite{viswanath-low-power-rtl-verification} and \cite{viswanath-dedicated-rewriting-verification}), the authors present \textit{dedicated rewriting}, a rewriting methodology to automatically prove the correctness of low power transformations at the RTL-level. They propose a highly automated deductive verification technique which is fine tuned for low power transformations. They prove the equivalence of two Verilog RTL designs, one derived from the other after the application of a low power transformation.

\subsection{Verification of Low Power features at Platform / System level}
\label{subsection:sys-level-verification}
 
 In order to accomplish this, typically companies use a combination of pre-silicon simulation, emulation techniques including complex FPGAs to emulate the entire chip/SoC RTL, and build platform level validation/verification tools that can include the ability to boot entire operating system on such FPGA systems. SoftSDV \cite{intel-softsdv} from Intel, for example, is a pre-silicon functional verification tool. However, this does not allow for detailed power estimation, modeling and verification. Several internal, proprietary (and costly) validation systems are used typically for validation of power management features. Viswanath et al. \cite{viswanath-rajeev-jolpe-survey} present a comprehensive of system level verification techniques. 
 
 Industrial designs rely heavily on ensuring that once the silicon arrives, power management can be validated as soon as possible, and thermal solutions can be built accurately for the specific form factors in consideration. In order to accomplish this, companies typically use FPGAs to emulate the SoC RTL, and build platform level validation/verification tools that can include the ability to boot entire operating system on such FPGA systems. Kapoor et al. \cite{soc-low-power-verification} present a good overview of the different techniques used in system level low power verification, the importance of using power intent specifications like UPF and simulation tools/methodologies that can accurately model power states/sequences. Mischkalla et al.  \cite{verification-virtual-prototyping} describe System-C based virtual prototyping techniques to perform power intent/sequence validation, and also propose using system level low power abstractions as possible extensions to UPF. This includes abstract definition of voltage relationships and dynamic aspects such as operating conditions. Muralidhar et al. \cite{dvcon-moorefield-pre-si-verification} discuss about HW-SW co-design and verifying energy efficiency features in pre-silicon, and the need for simulating end-to-end use cases in such verification methodologies. Targeted verification of each IP block, including CPU cores, GPUs, memory, and others can be done using traditional silicon verification techniques through a combination of random, targeted and functional PM tests. Since SOCs typically integrate third party IP blocks, specific PM related tests are needed for such IPs. Beyond the IPs, and going into the system level, a combination of different platforms and environments are used for different aspects of pre-silicon verification. These include Virtual Platforms (VP, where an entire OS can be booted quickly on a simulated system model), FPGAs (for specific hardware), Hybrid Virtual platforms (VP plus FPGA), System Level Emulation (SLE) platforms that is a complex FPGA that simulates parts of the chip or the entire chip. Each environment is best suited for a specific set/category of pre-silicon verification. Some of them can support production OS boot in reasonable times for SW development/co-design/debug. For thermal validation, different form factor devices are built early on and are analyzed in heat chambers. Based on the thermal hot spots, appropriate thermal control algorithms are defined and fine tuned. This is a costly, but accurate way of ensuring that thermal management on the devices are validated effectively. Usually, a multi-pronged strategy is used that could be a combination of all or some of these environments and techniques.
  

\section{Energy Efficiency Standards, Benchmarks and Cross Layer Energy Efficiency}	\label{section:energy-standards}


In this section, we will discuss important industry consortiums, standards, benchmarks and regulations for energy efficient and sustainable computing. 

\subsection{Consortiums}
\label{subsection:ee-consortiums}
The Green Grid \cite{greengrid} is a global consortium dedicated to advancing energy efficiency in data centers founded by many companies like AMD, Dell, HP, IBM, Intel, VMware and many others. The Green500 \cite{green500} list rates supercomputers by energy efficiency, encouraging a focus on efficiency (megaflops/watt) rather than absolute performance. The Energy Efficient HPC (EEHPC) \cite{eehpc} is a group that is focused on driving implementation of energy conservation measures and energy efficient design of HPC systems. The working groups cover several aspects of EE HPC systems - infrastructure, cooling, efficient power sources, systems architecture, energy aware job scheduling, specifications (Power API) and benchmarks. The key motivation for \textbf{Power API} is that achieving practical exascale computing will require massive increases in energy efficiency across hardware and software. With every generation of new hardware, more power measurement and control capabilities are exposed, with in-chip monitoring rapidly increasing as there are more sensors to track process, voltage, and temperature across the die \cite{in-chip-monitoring}. EEHPC's Power API is a portable API for power measurement and control; it provides multiple levels of abstractions, and allows algorithm designers to add power and energy efficiency to their optimization criteria at the system level like energy-aware scheduling. Finally, such systems may not be able to operate all components at full capability for a range of reasons including temperature, power delivery or battery limitations, thereby requiring software to make appropriate choices about how to allocate the available power budget given many, and sometimes conflicting considerations. 

\subsection{Benchmarks}
\label{subsection:ee-benchmarks}

The Transaction Processing Performance Council (TPC) Energy specification \cite{tpcc-energy} augments existing TPC benchmarks with energy metrics. The metric is calculated as the ratio of the energy consumed by all components of the benchmark system (typically measured in watts-seconds) to the total work completed (typically measured as a number of transactions). The benchmark system (system under test) includes servers, storage systems, and also network components like switches.

SPECpower \cite{spec-power} is perhaps the first industry standard benchmark that measures power consumption in relation to performance for server-class computers. The workload exercises the CPUs, caches, memory hierarchy and the scalability of shared memory processors (SMPs) as well as the implementations of the JVM (Java Virtual Machine), JIT (Just-In-Time) compiler, garbage collection, threads and some aspects of the operating system. Other benchmarks which measure energy efficiency include SPECweb, SPECvirt, and VMmark and EEMBC's ULPMark \cite{eembc-ulpmark}.

\subsection{Standards}
\label{subsection:ee-standards}
The Energy Star \cite{energy-star} program sets regulations around energy efficiency requirements for computer equipment, along with a tiered ranking system for approved products for mostly idle, and some active workloads. It is run by the U.S. Environmental Protection Agency and U.S. Department of Energy to promote energy efficiency across all categories of computing and electronic systems using different standardized methods.

\subsubsection{California Energy Commission (CEC)}
\label{subsubsection:cec}

The California Energy Commission's \cite{cec} goal is to lead the state to a 100 percent clean energy future. As the state's primary energy policy and planning agency, the Energy Commission plays a critical role in creating the energy system of the future. CEC has been driving some of the most stringent energy regulatory standards for computing systems and other electronic appliances via Energy Star and related programs that have now been adopted in different countries around the world.

\subsubsection{IEEE P2416 Standard for Power Modeling of Electronic Systems}
\label{subsubsection:ieee-p2416}
IEEE P2416 \cite{ieeep2416} defines a framework for the development of parameterized, accurate, efficient, and complete power models for hardware IP blocks and the entire system that can be used for power modeling and analysis. It is based on process, voltage, and temperature (PVT) independence and defines power and thermal management interfaces for hardware models and also workload and architecture parameterization. Such models are suitable for use in software development and hardware design flows, as well as for representing both pre-silicon estimates and post-silicon data. The working group recently released a version of this standard \cite{ieee-p2416-release}.
    
\subsubsection{IEEE P2415 Unified HW Abstraction and Layer For Energy Proportional Systems}
\label{subsubsection:ieee-p2415}
IEEE P2415 standard \cite{ieeep2415} intends to define the syntax and semantics for energy oriented description of hardware, software and is expected to be compatible with the IEEE 1801 (UPF) and IEEE P2416 standards to support an integrated flow across architecture, design, estimation and system software. The standard complements functional models in VHDL/Verilog/SystemVerilog/ SystemC by providing an abstraction of the design hierarchy and the design behavior with regard to power/energy usage in order to fill a key gap - current IEEE P1801 (UPF) is focused on the voltage distribution structure in design at RTL and below, has minimal abstraction for time, but depends on other hardware oriented standards to abstract events, scenarios, clock or power trees that are required for energy proportional design, verification, modeling and management of electronic systems. 

It is aimed at enabling specifying, modeling, verifying, designing, managing, testing and measuring the energy features of the device, covering both the pre- and post-silicon design flow. On the hardware side, the description aims to cover enumeration of components (SOC, board, device), memory map, bus structure, interrupt logic, clock and reset tree, operating states and points, state transitions, energy and power attributes; on the software side the description aims to cover software activities and events, scenarios, external influences (including user input) and operational constraints; and on the power management side the description aims to cover activity dependent energy control. This is quite an ambitious goal indeed, but a very important one for future energy proportional systems. The necessary abstractions of hardware, as well as layers and interfaces in software are not yet defined by any existing standards. This standard aims to address energy proportionality through tight interplay between energy-efficient hardware and energy-aware software. It provides new design, verification, modeling, management and testing abstractions and formats for hardware, software and systems to model energy proportionality, and enables the design methodology that naturally follows the top-down approach – from the system and software down to the hardware. 
Unfortunately, this standard seems to be inactive in recent times \cite{what-happened-to-upf}. 

\subsection{Cross Layer Optimizations for Energy Efficiency}
\subsubsection{Geo PM}
\label{subsubsection:geopm}
The Global Extensible Open Power Manager (GEOPM) (Eastep et al. \cite{geopm-hpc}) is an open source runtime framework with an extensible architecture enabling new energy management strategies in HPC systems. Different plugins can be tailored to the specific performance or energy efficiency priorities of each HPC center. It can be used to dynamically coordinate hardware settings across all compute nodes used by an application in response to the application's behavior and requests from the resource manager. The dynamic coordination is implemented as a hierarchical control system for scalable communication and decentralized control. The hierarchical control system can optimize for various objective functions including maximizing global application performance within a power bound or minimizing energy consumption. 
    
	
\subsubsection{Software-defined Power Meters: Power API, WattsKit}
\label{subsubsection:power-api}
Software-defined power meters are configurable software libraries that can estimate the power consumption of software in real-time. PowerAPI (Colmant et al.\cite{power-api}, \cite{colmant:powerapi}) and WattsKit \cite{colmant:wattskit} are some of the middleware toolkits for building software-defined power meters. PowerAPI takes an interesting approach to energy consumption measurements. It does not require any external device to measure energy consumption and is a purely software approach where the estimation is based on analytical models that characterize the consumption of various hardware components (CPU, memory, disk, etc.). PowerAPI is based on a highly modular architecture where each module represents a measurement unit for a specific hardware component. Power API is a novel toolkit that uses a learning technique to automatically learn the power model of a CPU, independently of the features and the complexity it exhibits. It automatically explores the space of hardware performance counters made available by a given CPU to isolate the ones that are best correlated to the power consumption of the host, and then infers a power model from the selected counters.

\section{The Road Ahead and New Trends}
\label{section:discussion}
The semiconductor industry has gone through several decades of evolution; compute performance has increased by orders of magnitude that was made possible by continued technology scaling, improved transistor performance, increased integration to realize novel architectures, extreme form factors, emerging workloads, and reducing energy consumed per logic operation to keep power and thermal dissipation within limits. We have worked around fundamental issues like ILP limits, end of Dennard scaling, and Amdahl's limit on multi-core performance. More recently, and expectedly, there has been a slowdown of Moore's Law. The following trends will continue to inexorably push computing beyond current limits:
	\begin{enumerate}
	    \item \textbf{Lower process nodes}: The industry is currently in the sub-10nm node, and a shift to 5nm and 3nm will provide a few generations of performance gains and energy efficiency, but requiring new transistor architectures like nanosheets and nanowires beyond today's FinFETs. Stacking nanosheets will provide perhaps the last step step in Moore's Law (Ye et al. \cite{ieee/nanosheet}). 
	    
	    \item \textbf{Heterogeneous architectures}: Mainstream computing will continue to see heterogeneous architectures comprising of CPUs, GPUs, domain specific accelerators and programmable hardware (FPGAs) across the spectrum with tightly integrated solutions.  
	    
	    \item \textbf{Exascale and beyond}: Research and industry will continue the push to build exascale systems using new architectures (Borkar et al.  \cite{Borkar-exascale}) and computing paradigms like mixing von Neumann and non-von Neumann models \cite{intelcsa}.
	    
	    \item \textbf{Sub-threshold voltage designs}: At the other end of the spectrum, sub-threshold and near-threshold voltage designs and techniques will enable ultra low power IoT and embedded/wearable markets that consume drastically lower power than traditional chips\cite{subthreshold/techniques}, \cite{subthreshold/design}.  Companies such as Ambiq Micro, PsiKick and Minima Processor, among others, have matured techniques developed in academia (Univ of Michigan, MIT and VTT Technical Research Center at Finland, respectively) to develop ultra low power chips that operate at 0.1-0.2 V range, with wide dynamic range as well, all the way up to 0.8 V \cite{sub-threshold-startups}.
	    
	    \item \textbf{Rise of non-x86 architectures and custom chips}: The last couple of years has also seen the rise of ARM architectures in enterprise systems (laptops), data center and cloud systems that were traditionally x86-based, which was primarily due to the unmatched performance of Intel and AMD processors. Recently, we have seen Apple's M1 chip \cite{apple/m1} in the personal PC domain and chips such as Amazon Web Services' Graviton2 \cite{aws/graviton2} for the data center and ARM in HPC \cite{hpc/arm}. We believe this trend of custom silicon will continue to push the boundaries of architectural innovation. 

	   \item \textbf{Energy efficient hardware}: We will see newer, open standards based (RISC-V, for eg.), energy-efficient architectures as computer architecture becomes more multi-disciplinary cross cutting computer science and cognitive science as our understanding of nature and the human mind evolves (neuromorphic and bio-inspired chips, for example). TinyML \cite{tinyml} is an important emerging area of machine learning under the 1mW power envelope. Similarly, software-defined hardware \cite{sw-defined-hw} is an important area of reconfigurable systems.
	    
	    \item \textbf{Energy-aware software}: Software and operating systems will need to evolve in lock-step fashion to utilize energy efficient hardware across different categories of systems and under varying energy efficiency/thermal constraints and challenges such as dark silicon and accelerator limits.
	    
	    \item \textbf{Cross Layer Energy Efficiency, Standards}: Systems will necessitate a tight interplay between energy efficient hardware and energy aware software through standardized cross layer abstractions across architecture, design, modeling and simulation, implementation, verification and optimization of complete systems.
	    
	    \item \textbf{Domain-specific stacks}: Across different computing domains (ultra low power/IoT, edge, mainstream, cloud, HPC and exascale), the industry will see highly optimized domain-specific stacks that are built using modular, standardized hardware-software interfaces and components. For example, Tesla's full self-driving solution (FSD) (Talpes et al. \cite{tesla-fsd}), which is a tightly integrated, domain-specific system for autonomous driving with a TDP of under 40W.
	    
	    \item \textbf{Neuromorphic Computing and other non-von Neumann systems}: Several industrial systems are available now that implement non-von Neumann architectural and programming models such as IBM’s TrueNorth \cite{ibmtruenorth} and Intel's Loihi (Davies et al. \cite{intelloihi}). Both are based on Spiking Neural Networks to demonstrate neuromorphic architectures. Poihiki Springs took this further with a rack-mounted chassis enclosing 768 Loihi chips, FPGA interface boards, and an integrated IA host CPU. Davies et al. \cite{intel/loihi/survey} describes the latest results of how specific types of deep learning algorithms (brain-inspired networks) perform with orders of magnitude lower latency and energy. Such architectures and systems will continue to push the boundaries of non-von Neumann computing.
	    
	    \item \textbf{Quantum Computing}: Quantum computing is on the horizon now, with several experimental quantum computing architectures being built and used for specific optimization problems. Amazon Web Services provides Braket, a fully managed cloud service that allows scientists, researchers, and developers to begin experimenting with computers from multiple quantum hardware providers in a single place \cite{aws/braket}. Similarly, Microsoft provides a quantum development kit for the Q\# quantum programming language and Azure Quantum hardware based on topological quantum computing. Intel is aiming for "quantum practicality" \cite{intel/quantum} with its attempt to use silicon spin qubits that look exactly like a transistor; this would enable high volume fabrication for silicon-based quantum computing. For the foreseeable future, quantum computers will at best be accelerators that will interface with classical computers \cite{qc-delft}. Getting such systems to work is the immediate focus across research and industry.
	    
	    \item \textbf{Thermodynamic computing}: As we push the boundaries of computing and look at how to make computers function more efficiently, researchers are probing the foundations of \textit{thermodynamic computing} \cite{thermodynamic-computing} based on the observation that thermodynamics drives the self-organization and evolution of natural systems and, therefore, thermodynamics might drive the self-organization and evolution of future computing systems, making them more capable, more robust, and highly energy efficient. It is not very clear what thermodynamic computing will look like at this time of writing. 
	\end{enumerate}
	
	\section{Summary and Conclusions}
	\label{section:summary}
	Computing systems have undergone a tremendous change in the last few decades with several inflexion points. While Moore's law guided the semiconductor industry to cram more and more transistors and logic into the same volume, the limits of instruction-level parallelism (ILP) and the end of Dennard's scaling drove the industry towards multi-core chips; we have now entered the era of domain-specific architectures, pushing beyond the memory wall. However, challenges of dark silicon and other limits will continue to impose constraints. Overall energy efficiency encompasses multiple domains - hardware, SOC, firmware, device drivers, operating system runtime and software applications/algorithms and therefore must be done at the entire platform level in a holistic way and across all phases of system development. 
	
	This survey brings together different aspects of energy efficient systems, through a systematic categorization of \textit{specification}, \textit{modeling} and \textit{simulation}, \textit{energy efficiency techniques},  \textit{verification}, \textit {energy efficiency benchmarks, standards, consortiums and cross layer efforts} that are crucial for next generation computing systems. Future energy efficient systems will need to look at all these aspects holistically, through cross-domain, cross-layer boundaries and bring together energy efficient hardware and energy-aware software.
		
	Trends indicate that systems will continue to evolve, pushing the boundaries of technology, architecture, design and manufacturing. For future systems, the power wall will be the boundary condition around which computing systems will evolve across the ends of the computing spectrum (ultra low power devices to large HPC/exascale systems), through a tight interplay between energy efficient hardware and energy-aware software.

	\bibliographystyle{ACM-Reference-Format}
    \bibliography{eesurvey}
	
	
	%

\end{document}